\pdfoutput=1
\documentclass[twocolumn]{aastex63}

\usepackage{epstopdf}
\usepackage{lineno}
\usepackage{xspace}
\usepackage{graphicx}
\usepackage{amssymb}
\usepackage{amsmath}
\usepackage{natbib}
\usepackage{float}
\usepackage{hyperref}

\shorttitle{MIR Diagnostics of Young Radio Galaxies}
\shortauthors{Kosmaczewski et al.}

\begin{document}

\title{Mid-Infrared Diagnostics of the Circumnuclear Environments\\ of the Youngest Radio Galaxies}

\correspondingauthor{E.~Kosmaczewski}
\email{e.kosmaczewski@oa.uj.edu.pl}

\author{E.~Kosmaczewski}
\affiliation{Astronomical Observatory of the Jagiellonian University, ul. Orla 171, 30-244 Krak\'ow, Poland}

\author{\L .~Stawarz}
\affiliation{Astronomical Observatory of the Jagiellonian University, ul. Orla 171, 30-244 Krak\'ow, Poland}

\author{A.~Siemiginowska}
\affiliation{Harvard Smithsonian Center for Astrophysics, 60 Garden Street, Cambridge, MA 02138, USA}

\author{C.~C.~Cheung}
\affiliation{Naval Research Laboratory, Washington, DC 20375, USA}

\author{L.~Ostorero}
\affiliation{Dipartimento di Fisica -- Universit\`a degli Studi di Torino and Istituto Nazionale di Fisica Nucleare (INFN), Via P. Giuria 1, I-10125 Torino, Italy}

\author{M.~Sobolewska},
\affiliation{Harvard Smithsonian Center for Astrophysics, 60 Garden Street, Cambridge, MA 02138, USA}

\author{D.~Kozie{\l}-Wierzbowska}
\affiliation{Astronomical Observatory of the Jagiellonian University, ul. Orla 171, 30-244 Krak\'ow, Poland}

\author{A.~W\'ojtowicz}
\affiliation{Astronomical Observatory of the Jagiellonian University, ul. Orla 171, 30-244 Krak\'ow, Poland}

\author{V.~Marchenko}
\affiliation{Astronomical Observatory of the Jagiellonian University, ul. Orla 171, 30-244 Krak\'ow, Poland}

\begin{abstract}
We present a systematic analysis of the mid-infrared (MIR) properties of the youngest radio galaxies, based on low-resolution data provided by the {\it WISE} and {\it IRAS} satellites. We restrict our analysis to sources with available X-ray data that constitute the earliest phase of radio galaxy evolution, i.e. those classified as Gigahertz Peaked Spectrum (GPS) and/or Compact Symmetric Objects (CSOs). In our sample of 29 objects, we find that the host galaxies are predominantly red/yellow ellipticals, with some of them displaying distorted morphology. We find a variety of MIR colors, and observe that the sources in which the MIR emission is dominated by the ISM component uniformly populate the region occupied by galaxies with a wide range of pronounced ($\geq 0.5 M_{\odot}$\,yr$^{-1}$) star formation activity. We compare the MIR color distribution in our sample to that in the general population of local AGN, in the population of evolved FR\,II radio galaxies, and also in the population of radio galaxies with recurrent jet activity. We conclude that the triggering of radio jets in AGN does not differentiate between elliptical hosts with substantially different fractions of young stars; instead there is a relationship between the jet duty cycle and the ongoing star formation. The distribution of the sub-sample of our sources with $z<0.4$ on the low-resolution MIR vs. absorption-corrected X-ray luminosity plane is consistent with the distribution of a sample of local AGN. Finally, we comment on the star formation rates of the two $\gamma$-ray detected sources in our sample, 1146+596 \& 1718--649.
\end{abstract}


\section{Introduction} \label{sec:intro}

Compact radio sources constitute a particularly interesting, yet diverse, class of active galactic nuclei (AGN), with newly born radio structures (jets and lobes) fully confined within their host galaxies \cite[see][for a review]{Odea98}. As such, they provide direct insight into the mechanisms that lead to triggering the production of relativistic jets in AGN \cite[e.g.,][]{Czerny16}, as well as insight into the complex dynamics of feedback between evolving supermassive black holes and interstellar medium \cite[e.g.,][]{Tadhunter16,Wagner16}. 

To study these dynamics, specific evolutionary models for compact radio galaxies have been developed \cite[e.g.,][]{Kawakatu08}, and certain predictions have been made regarding their multi-wavelength emission and absorption properties \cite[e.g.,][]{Stawarz08}. These models have been tested against various multi-frequency datasets \citep{Ostorero10,Migliori14}. There, however, still remains a lack of consensus on the relative contribution from jets and lobes, accretion disks and circum-nuclear dusty tori, or, to some extent, the interstellar medium (ISM) itself, to the observed radiative output of such systems at infrared, X-ray, and $\gamma$-ray photon energies.

In this context, we focus on a systematic investigation of the mid-infrared (MIR) properties of the most compact radio galaxies. These sources are confirmed spectroscopically in the radio domain as `GHz-peaked spectrum' (GPS) sources, and/or morphologically (using high-resolution radio interferometers) as `compact symmetric objects' (CSOs). In order to study the MIR properties, we utilize data provided by the Wide-field Infrared Survey Explorer ({\it WISE}), as well as archival data from the Infrared Astronomical Satellite ({\it IRAS}) augmented in a few cases by {\it Spitzer} Space Telescope observations. We restrict our sample only to those objects which have been observed in X-rays with either the XMM-{\it Newton} or the  {\it Chandra} X-ray Observatory, so that multi-wavelength diagnostics can be applied to disclose the origins of the observed fluxes.

Previously, the infrared emission of GPS/CSOs had been analyzed by \cite{Heckman94} in the mid-far infrared domain, based on low-resolution data from {\it IRAS}, and by \cite{Fanti00} in the far infared, using the Infrared Space Observatory (ISO) data. These authors argued that the mid-to-far infrared emission of compact radio galaxies is comparable to that of the extended radio galaxies with matching radio powers and redshifts. Therefore, these sources are most likely dominated by an AGN component, the hot dusty tori in particular. 

More recently, \cite{Willett10} presented the MIR observations of eight nearby CSOs (redshifts $z \lesssim 0.1$) from {\it Spitzer}, finding a diversity in their spectral properties. An overwhelming majority of the sources in their sample displayed a contribution from both circum-nuclear dust, heated by the central AGN, and star formation activity within the host galaxy. \cite{Willett10} argued, moreover, that the star formation rates in the studied sources, estimated at the level of $(0.3-50) M_{\odot}$\,yr$^{-1}$ based on the detected polycyclic aromatic hydrocarbon (PAH) emission, implies a close link between the triggering of radio jets and galaxy mergers. On the other hand, \cite{Tadhunter11} and \cite{Dicken12} proposed that the enhanced star formation activity in compact radio sources over that observed in evolved radio galaxies, indeed seen in their flux limited samples, may be instead due to an observational selection effect. This follows from the enhanced radiative efficiency of compact radio-emitting jets and lobes in the former class of objects \cite[see also][for the discussion]{Odea16}.

In the X-ray domain, the first dedicated studies of GPS/CSOs with the high-angular resolution telescopes XMM-{\it Newton} and {\it Chandra}, have been reported by \cite{Guainazzi04,Guainazzi06,Vink06,Siemiginowska08,Siemiginowska09, Siemiginowska16,Tengstrand09}, and \cite{Sobolewska19a,Sobolewska19b}. Since the radio structures of the targeted sources remain unresolved even on arcsec scales, various origins of the detected fluxes within the 0.1--10\,keV range have been considered by these authors including: accretion disk coronal emission, jet/lobe non-thermal radiation, as well as emission of hot gas within the ISM, modified (shocked) by expanding compact radio structures. 

In most cases, unfortunately, rather limited photon statistics preclude any more in-depth spectral modeling, which would enable a robust discrimination between various emission models. The X-ray continua of compact radio galaxies were typically found to be consistent with a single power-law, with photon indices $\Gamma \sim 1-2$, moderated by hydrogen column densities $N_{\rm H} \lesssim 10^{22}$\,cm$^{-2}$. Such spectra could be accommodated by either the disk coronal emission scenario, with modest intrinsic absorption, or the model in which observed X-ray fluxes are dominated by inverse-Comptonization of various ambient photon fields (and the infrared emission of circumnuclear dusty tori in particular) within compact radio lobes shining through the ISM of host galaxies \citep{Ostorero10}. Only in a few cases, subjected to deeper exposures (1404+286, 1511+0518 and 2021+614), was there a detection of the neutral fluorescence iron line and/or hydrogen column densities in excess of $10^{23}$\,cm$^{-2}$, indicating the disk coronal emission is seen through a Compton-thick circumnuclear tori \citep{Guainazzi04,Siemiginowska16,Sobolewska19a,Sobolewska19b}. On the other hand, for the particular case of 1718--649, the observed X-ray emission appeared contributed to by a hot, collisionally ionized gas located within the central parts of the host galaxy \citep{Siemiginowska16,Beuchert18}. 

Below, in Section\,\ref{sec:sample} we discuss the selection of the sample including the most compact radio galaxies observed in X-rays. In Section\,\ref{sec:data} we describe the MIR data acquisition for the selected sources, specifically the {\it WISE}, {\it IRAS}, and {\it Spitzer} archival observations. In Section\,\ref{sec:results} we discuss the results of our analysis, and our main conclusions are outlined in Section\,\ref{sec:conclusions}. Throughout the paper we assume modern $\Lambda$CDM cosmology with $H_{0}=70$\,km\,s$^{-1}$\,Mpc$^{-1}$, $\Omega _{\rm m}=0.3$, and $\Omega _{\Lambda}=0.7$.
 
 \begin{turnpage}
\begin{deluxetable*}{crrrcrrcccccc}[t!]
\movetableright=0.05cm
\tabletypesize{\scriptsize}
\tablecaption{MIR properties of X-ray detected CSOs}
\tablewidth{0pt}
\tablehead{
\colhead{Name} & \colhead{$z$} & \colhead{$d_{\rm L}$} & \colhead{LS} & \colhead{class}&  \colhead{W1--W2} & \colhead{W2--W3} & \colhead{{\it WISE}} & \colhead{IRAS $L_{\rm 12\,\mu m}$} & \colhead{MIPS $L_{\rm 24\,\mu m}$} & \colhead{WISE $L_{\rm 12\,\mu m}$} & \colhead{$L_{\rm 2-10\,keV}$} & \colhead{Ref.} \\
\colhead{~~} &  \colhead{~~} & \colhead{[Mpc]} & \colhead{[pc]}&  \colhead{~~} &\colhead{[mag]} & \colhead{[mag]} & \colhead{color} & \colhead{[$10^{43}$\,erg/s]} & \colhead{[$10^{43}$\,erg/s]} & \colhead{[$10^{43}$\,erg/s]} & \colhead{[$10^{43}$\,erg/s]} & \colhead{~~} \\
\colhead{~~} & \colhead{~~} & \colhead{~~}  & \colhead{~~} &  \colhead{~~}& \colhead{~~} & \colhead{~~} &  \colhead{~~} & \colhead{~~} & \colhead{~~} & \colhead{~~} & \colhead{~~} & \colhead{~~} \\
\colhead{(1)} & \colhead{(2)} & \colhead{(3)} & \colhead{(4)} & \colhead{(5)} &  \colhead{(6)} & \colhead{(7)} & \colhead{(8)} & \colhead{(9)} & \colhead{(10)} & \colhead{(11)} & \colhead{(12)}& \colhead{(13)}
}
\tiny
\startdata
0019--000 & 0.305 & 1521 & 220 & GPS & 0.472 & 2.441 & G & -- & -- & 3.53 $\pm$ 0.11 & 0.55$^{\wedge}$  & T09\\
0026+346 & 0.517 & 2852 & 190 & GPS &0.743 & 2.491 & Un & -- & -- & 21 $\pm$ 0.3 & 23 $\pm$ 2 & G06\\
0035+227 & 0.096 & 418 & 21.8 & CSO & 0.148 & 1.265 & G & -- & -- & 0.425 $\pm$ 0.007 & 0.075 $\pm$ 0.034 & S16\\
0108+388 & 0.669 & 3907 & 22.7 & CSO & 0.132 & 3.159 & G & 9.86 $\pm$ 9.16 & 3.79 $\pm$ 1.40 & $<$13.1  & 7 $\pm$ 3  & T09,S16\\
0116+319 & 0.059 & 255& 70.1 & CSO & 0.035 & 1.682 & S$^{\star}$  & 0.595 $\pm$ 0.056 & 0.609 $\pm$ 0.046 & 0.629 $\pm$ 0.003 & $<$0.10$^{\dagger}$ & S16\\
0402+379 & 0.055 & 234 & 7.3 &CSO & 0.099 & 1.721 & G & 0.48 $\pm$ 0.07 & 0.949 $\pm$ 0.059 & 0.608 $\pm$ 0.004 & 0.041 $\pm$ 0.016 & R14\\
0428+205 & 0.219 & 1044 & 653 &  GPS &0.323 & 2.648 & G & -- & --& 3.93 $\pm$ 0.08 & 1.4 $\pm$ 0.6 & T09\\
0500+019 & 0.585 & 3319 & 55 &  GPS &1.047 & 2.892 & Q & 63.4 $\pm$ 6.5 & 87.2 $\pm$ 9.8 & 59.3 $\pm$ 0.5 & 50 $\pm$ 6 & T09\\
0710+439 & 0.518 & 2868 & 87.7  &  CSO &0.726 & 2.997 & Sy & 43.8 $\pm$ 4.4 & 107 $\pm$ 7 & 43.4 $\pm$ 0.4 & 39.40 $\pm$ 3.15 & T09,S16\\
0941--080 & 0.228 & 1100& 148 & GPS &0.388 &  2.444 & G* & -- & -- & 2.57 $\pm$ 0.05 & 0.091 $\pm$ 0.075 & T09\\
1031+567 & 0.460 & 2480& 109 & CSO & 0.999 & 2.922 & Q & 17.6 $\pm$ 3.0 & 16.5 $\pm$ 4.7 & 15.6 $\pm$ 0.3 & 2.2 $\pm$ 0.2  & T09,S16\\
1117+146 & 0.362 & 1874& 306 &GPS & 0.406 &  3.599 & SB & -- & -- & $<$5.87  & 1.40 $\pm$ 0.19 & T09\\
1146+596 & 0.011 & 47& 933$^{\diamond}$  & CSO$^{\bullet}$ & --0.011 & 1.237 & G & 0.0816 $\pm$ 0.0022 & 0.0818 $\pm$ 0.0033 & 0.1010 $\pm$ 0.0002 & 0.007  & U05\\
1245+676 & 0.107 & 478& 9.6 &CSO & 0.130 & 1.409 & G & 0.33 $\pm$ 0.14 & 0.358 $\pm$ 0.098 & 0.531 $\pm$ 0.006 & 0.031$^{\dagger\dagger}$ & W09,S16\\
1323+321 & 0.368 & 1908& 247 &GPS & 0.303 & 2.014 & G & -- & -- & 2.54 $\pm$ 0.11 & 3.7 $\pm$ 0.4 & T09\\
1345+125 & 0.122 & 551& 166  &CSO & 1.308 & 3.930 & Sy$^{\star}$ & 83.3 $\pm$ 1.7 & 216 $\pm$ 4 & 87.9 $\pm$ 0.2 & 7.8 $\pm$ 1.5 & T09,J13\\
1358+624 & 0.431 & 2298& 218 & GPS &1.210 & 2.592 & Q$^{\star}$ & 31.0 $\pm$ 2.4 & 37.8 $\pm$ 2.8 & 35.4 $\pm$ 0.2 & 30 $\pm$ 20 & T09\\
1404+286 & 0.077 & 336& 10.0 & CSO & 1.018 & 3.063 & Q & 50.9 $\pm$ 0.9 & 65.2 $\pm$ 1.3 & 51.6 $\pm$ 0.1 & 0.45 $\pm$ 0.06$^{\bf C}$ & T09,S16,S19b\\
1511+0518 & 0.084 & 370& 7.3 & CSO & 1.233 & 2.899 & Q & -- & 51.1 $\pm$ 3 & 52.2 $\pm$ 0.1 & 3$^{\bf C}$ & S16\\
1607+268 & 0.473 & 2569& 240 & CSO & 0.287 & 2.679 & S$^{\star}$ & 4.21 $\pm$ 3.17 & 21.6 $\pm$ 4.4 & $<$7.27  & 3.79 $\pm$ 0.87 & T09,S16\\
1718--649 & 0.014 & 60.4& 2.0 & CSO & 0.136 & 2.462 & G & 0.322 $\pm$ 0.009 & 0.32 $\pm$ 0.01 & 0.401 $\pm$ 0.001 & 0.0154 $\pm$ 0.0024 & S16\\
1843+356 & 0.763 & 4612& 22.3 & CSO & 1.160 & 4.047 & Sy & -- & -- & 190 $\pm$ 1 & 5.60 $\pm$ 2.28 & S16\\
1934--638 & 0.183 & 845& 85.1 & CSO &0.609 & 3.360 & Sy* & 5.89 $\pm$ 0.62 & 19 $\pm$ 1 & 6.52 $\pm$ 0.03 & 0.60 $\pm$ 0.14 & S16,S19a\\
1943+546 & 0.263 & 1285& 107.1 & CSO & 0.628 & 2.191 & Un & -- & -- & 5.64 $\pm$ 0.04 & 0.73 $\pm$ 0.28 & S16\\
1946+708 & 0.101 & 444& 39.4 & CSO & 0.663 & 2.336 & Un & 1.75 $\pm$ 0.15 & 1.67 $\pm$ 0.10 & 1.69 $\pm$ 0.01 & 1.20 $\pm$ 0.18 & S16,S19a\\
2008--068 & 0.547 & 3056& 218 & CSO & 0.334 & 2.159 & G & -- & 7.68 $\pm$ 0.35 & 9.1 $\pm$ 0.4 & 6 $\pm$ 2 & T09\\
2021+614 & 0.227 & 1086& 16.1 & CSO & 1.287 & 3.009 & Q & $<$210  & -- & 64.9 $\pm$ 0.2 & 11.2$^{\bf C}$ & S16,S19a\\
2128+048 & 0.990 & 6364& 218 & GPS &0.688 & 3.352 & Sy & -- & 63.9 $\pm$ 2.3 & $<$52.4  & 29 $\pm$ 4 & T09\\
2352+495 & 0.238 & 1143& 117.3 & CSO & 0.688 & 2.714 & Sy & -- & -- & 4.25 $\pm$ 0.04 & 1.3$\pm$ 0.3  & T09,S16\\
\enddata
\tablenotetext{}{\\ {\bf col(1)} --- name of the source; 
{\bf col(2)} --- redshift; 
{\bf col(3)} --- luminosity distance;
{\bf col(4)} --- linear size of the radio lobes taken from the references provided in col(13), $^{\diamond}$\,except of 1146+596 cited from \citet{Perlman01}; 
{\bf col(5)} --- morphological/spectral radio classification: Compact Symmetric Object (CSO) based on the radio morphology, Gigahertz Peaked Spectrum (GPS) based on the spectral classification; classification follows the references given in col(13), $^{\bullet}$\,except of 1146+596 following from \citet{Perlman01}; 
{\bf col(6)} --- {\it WISE} Difference in the W1 (3.4\,$\mu$m) and W2 (4.6\,$\mu$m) color bands; 
{\bf col(7)} --- {\it WISE} Difference in the W2 (4.6\,$\mu$m) and W3 (12\,$\mu$m) color bands; 
{\bf col(8)} --- color classification: ``Galaxy'' (G), ``Starburst'' (SB), Quasar/Seyfert (Q/Sy), ``Uncertain'' (Un); possible contamination of {\it WISE} fluxes by background/foreground sources denoted by $^{\star}$; 
{\bf col(9)} --- {\it IRAS} 12\,$\mu$m luminosity;
{\bf col(10)} --- {\it Spitzer}/MIPS 24\,$\mu$m luminosity; 
{\bf col(11)} --- {\it WISE} 12\,$\mu$m luminosity calculated from the W3 band magnitude; 
{\bf col(12)} --- unabsorbed $2-10$\,keV luminosity taken from the references provided in col(13); $^{\wedge}$ extrapolated from the detection in the $0.5-2$\,keV band with photon index $= 1.61$; $^{\dagger}$\,estimated from the $0.5-2.0$ keV upper limit $<0.5\times 10^{-14}$\,erg\,cm$^{-2}$\,s$^{-1}$ assuming photon index $=2.0$; $^{\dagger\dagger}$\,calculated based on the $4.5-12.0$\,keV PN flux; $^{\bf C}$\,value corresponding to the ``Compton-thick'' scenario; 
{\bf col(13)} --- X-ray references: \citet{Guainazzi06} [G06], \citet{Jia13} [J13], \citet{Romani14} [R14], \citet{Siemiginowska16} [S16], \citet{Sobolewska19a} [S19a], \citet{Sobolewska19b} [S19b],  \citet{Tengstrand09} [T09], \citet{Ueda05} [U05], \citet{Watson09} [W09].
}
\label{table:1}
\end{deluxetable*}
\clearpage
\end{turnpage}

\section{Sample Selection} \label{sec:sample}

As mentioned in Section\,\ref{sec:intro} above, in this paper we consider only radio galaxies (i) for which the redshifts are measured, (ii) are characterized by compact radio structures with linear sizes LS\,$<1$\,kpc, and are classified morphologically as CSOs \emph{and/or} spectrally as GPS sources, and (iii) that have available X-ray data. After a careful inspection of literature studies, we construct a ``master sample'' of such objects, the majority of which make up the samples discussed by \citet{Tengstrand09} and \citet{Siemiginowska16}, with the addition of 0026+346, 0402+379, and 1146+596 \citep[see][]{Guainazzi06,Willett10,Ostorero17}. This amounts to a total of 29 targets, including 27 that have been detected in X-rays by either XMM-{\it Newton} or {\it Chandra}, one source with the X-ray flux measured by ASCA \citep[1146+596; see][]{Ueda05}\footnote{The analysis of the archival and previously unpublished {\it Chandra} data for 1146+596 will be presented in a forthcoming paper.}, and one source having a {\it Chandra} upper limit for its $0.5-2.0$\,keV flux \citep[0116+319; see][]{Siemiginowska16}. Table\,\ref{table:1} summarizes the basic information for the given sample, in particular, the $2-10$\,keV and MIR luminosities.

In the accompanying paper by \citet{Wojtowicz19}, we considered a smaller sample of the youngest X-ray detected radio galaxies, by imposing a much more restrictive selection criteria, namely (i) measured redshifts, (ii) strictly CSO morphological classification, in addition with \emph{measured kinematic ages}, (iii) core X-ray fluxes measured with either XMM-{\it Newton} or {\it Chandra}. All of the 17 sources analyzed by \citeauthor{Wojtowicz19} are included in our master sample as well. 

On the other hand, in our master sample we do not include X-ray detected compact radio galaxies lacking any robust GPS or CSO classification \citep[e.g., 0046+316, 1146+531, 1217+295, 1231+481, 1254+571; see][]{Sobolewska19a}, or classified before as GPS/CSOs but more recently re-classified as blazars \citep[e.g., 1413+135; see][]{Willett10}. We also do not discuss here the more evolved radio galaxies, classified morphologically as Medium Symmetric Objects (MSOs), and/or spectrally as Compact Steep Spectrum (CSS) sources \citep[see][]{Kunert14}. Finally, we do not include the Fanaroff-Riley type 0 (FR0) radio galaxies \citep{Glowacki17,Torresi18}.

All in all, our master sample of the youngest radio galaxies, with available X-ray data, is composed predominantly (17/29) from the flux- and volume-limited set of targets extracted by \citet{Tengstrand09} and \citet{Guainazzi06}, from a complete sample of GPS sources compiled by \citet{Stanghellini98}, by applying cuts on redshift, $z < 1$, and the 5\,GHz flux density,$F_{\rm 5\,GHz} > 1$\,Jy. For this sub-sample, the X-ray detection rate is particularly high, namely $100\%$ (including 0019-000 which has been detected with XMM-{\it Newton} only in the soft $0.5-2.0$\,keV band). Secondly, our master sample includes a set (10/29) of CSOs at $z<1$, \emph{with available kinematic age measurements}, selected for {\it Chandra} study by \citet{Siemiginowska16}. These sources are not flux or volume limited, and the radio luminosities are on average lower than that of the \citet{Tengstrand09} sub-sample; still, all but one (0116+319) of the sources selected by \citet{Siemiginowska16} have been detected with {\it Chandra}, despite short exposures, $\sim 5$\,ksec, for the majority of the targets. In addition to the sources in the above-mentioned sub-samples, we have also included two particularly compact and nearby CSOs, namely 0402+379 \citep[see][]{Willett10}, and 1146+596 \citep[see][]{Ostorero17}, in our master list. 

\begin{figure}[t!]
\centering
\includegraphics[width=0.475\columnwidth]{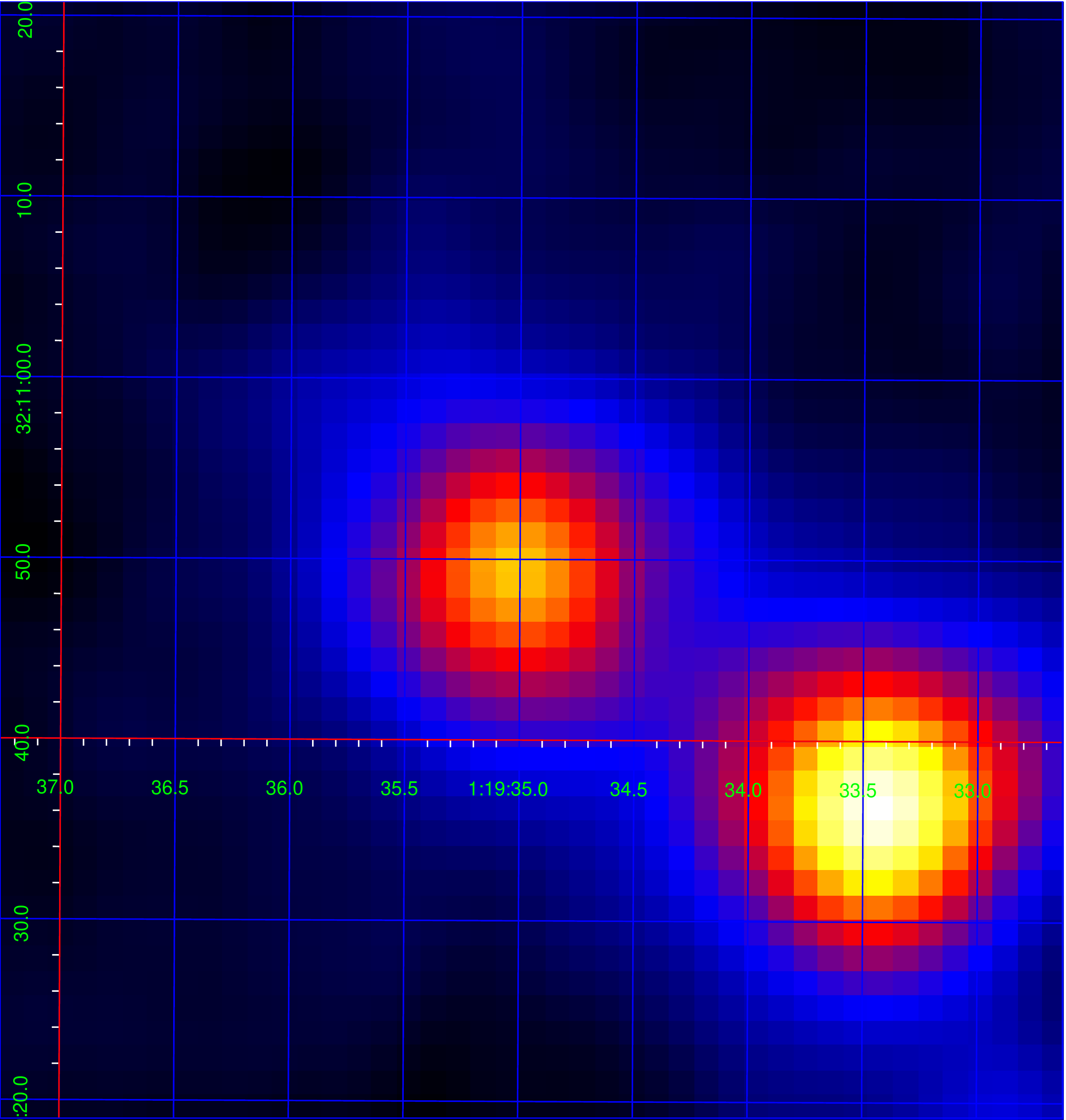}
\includegraphics[width=0.5\columnwidth]{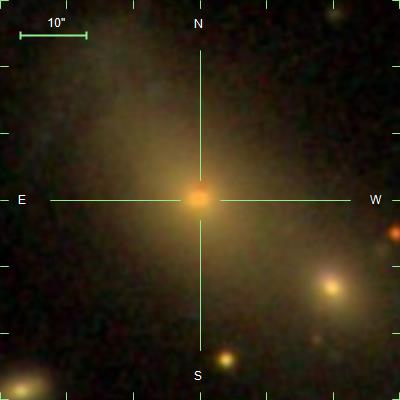} \\
\includegraphics[width=0.475\columnwidth]{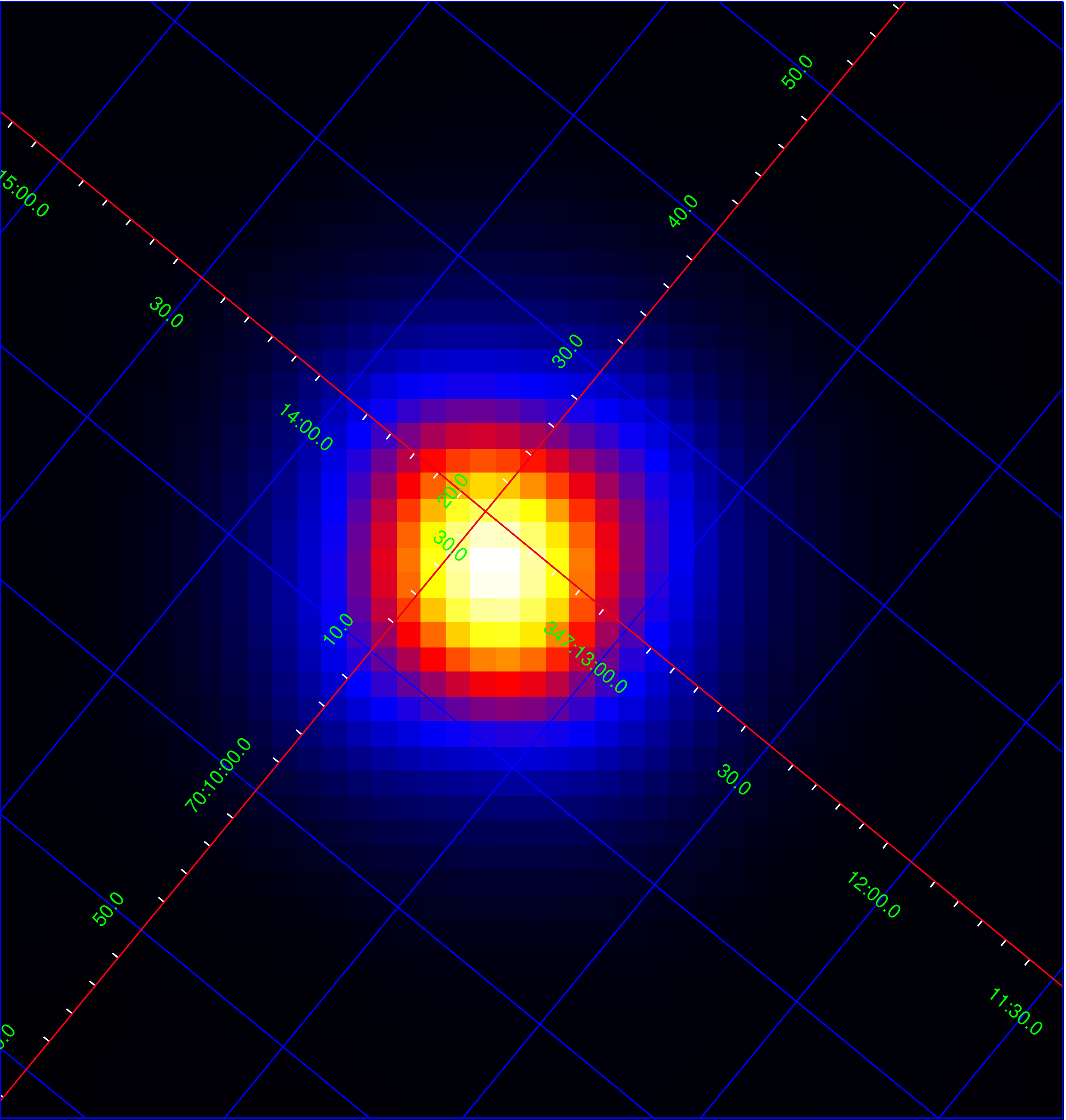} 
\includegraphics[width=0.5\columnwidth]{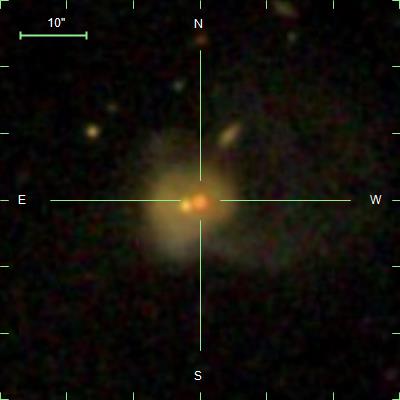} \\
\includegraphics[width=0.475\columnwidth]{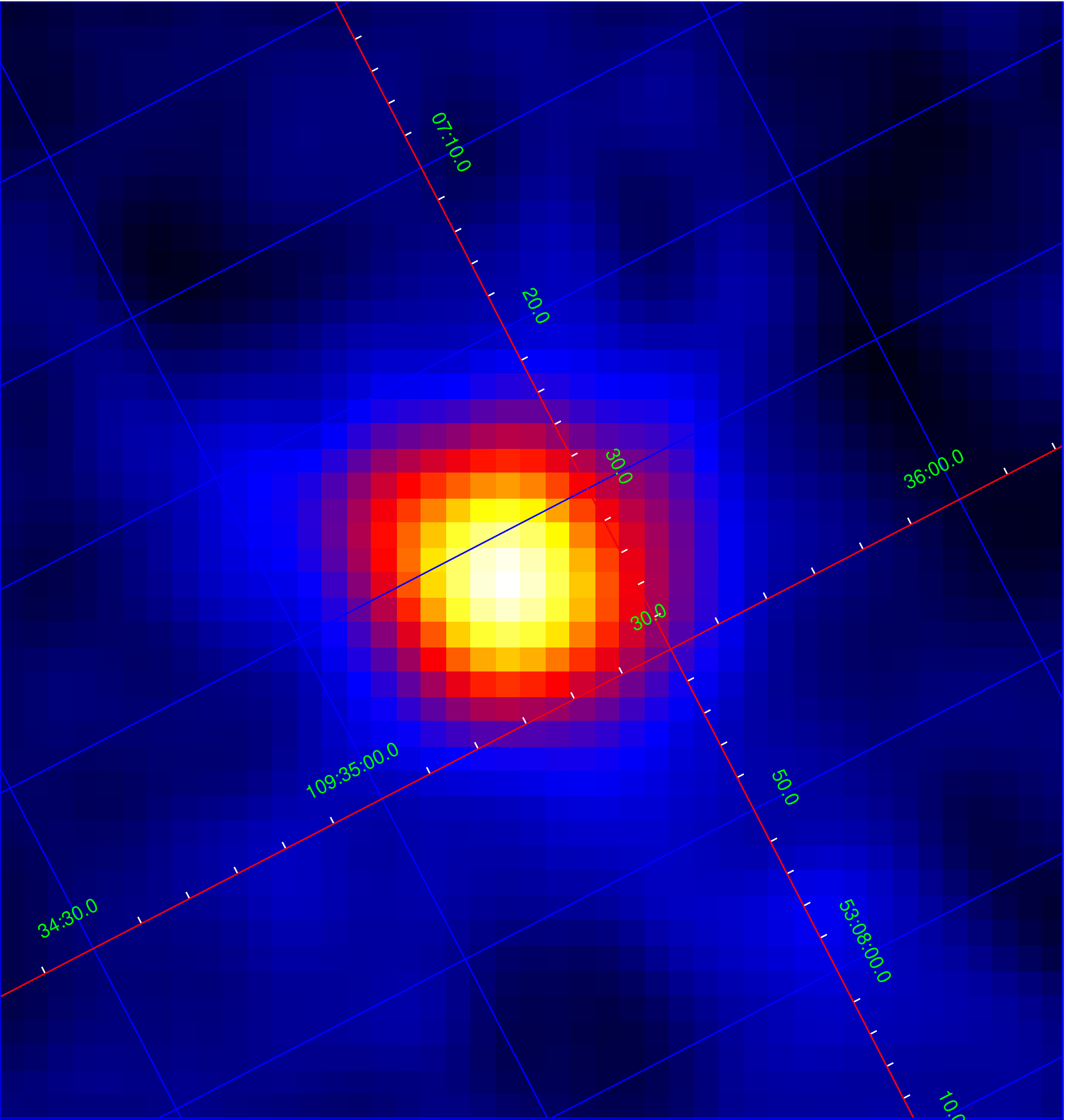} 
\includegraphics[width=0.5\columnwidth]{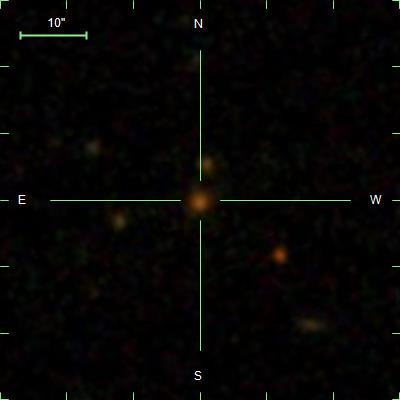} \\
\includegraphics[width=0.31\columnwidth]{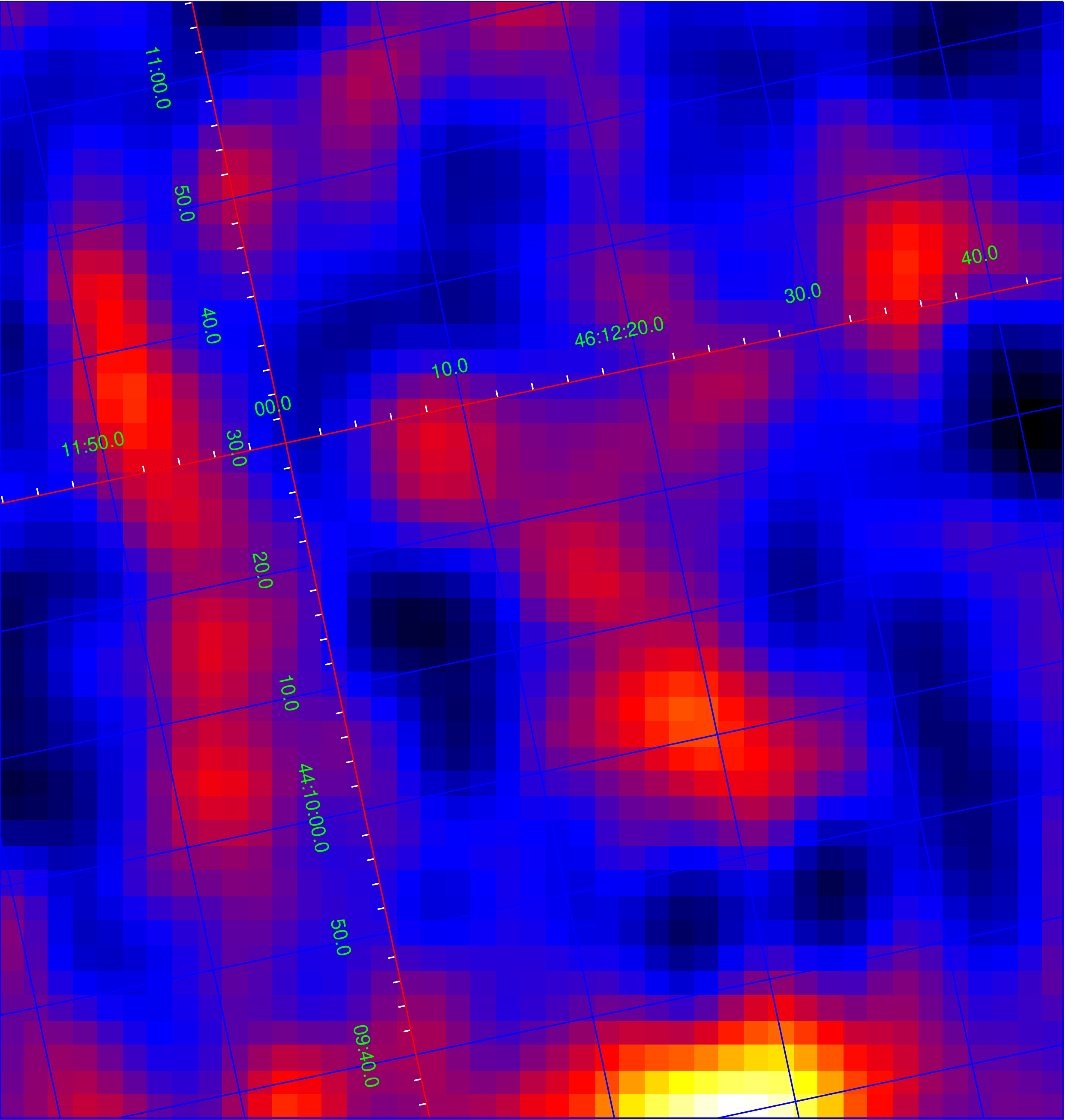} 
\includegraphics[width=0.31\columnwidth]{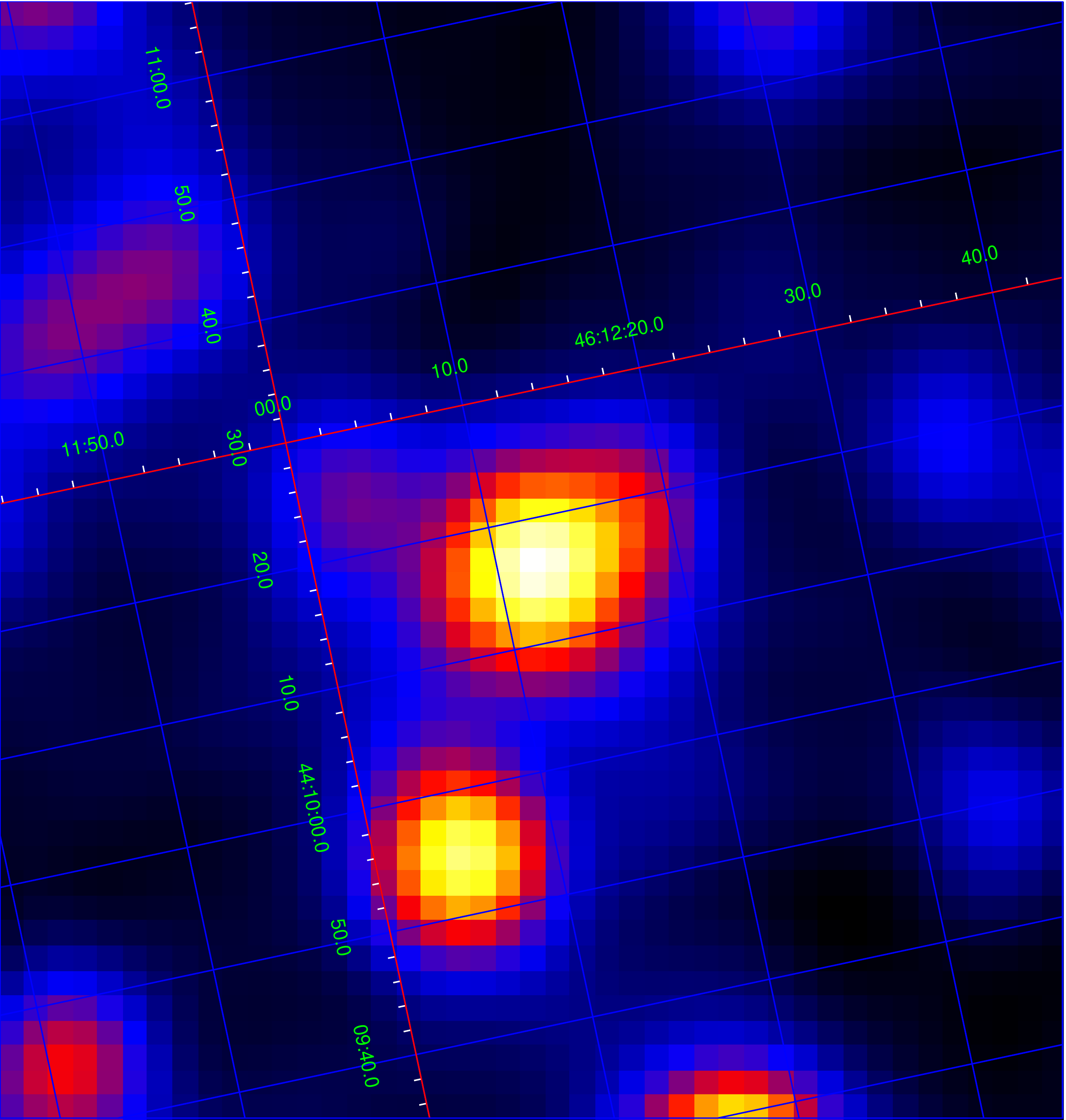} 
\includegraphics[width=0.31\columnwidth]{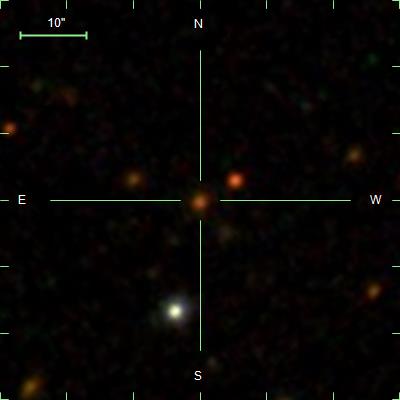} \\
\caption{{\it WISE} 12\,$\mu$m images with overlaid ecliptic coordinate grids (left panels), along with the SDSS gri color composite images (right panels) for $0116+319$, $1345 +125$, $1358+642$, and $1607+268$ (rows from top to bottom). In the case of $1607+268$ (bottom row), the  {\it WISE} 3.4\,$\mu$m image is presented in addition(central panel). Images are equal in size, being a square $60''$.}
\label{fig:contamination}
\end{figure}

\section{MIR Data Acquisition} \label{sec:data}
\subsection{MIR Data}

In this study we utilize three infrared telescopes: the Wide-field Infrared Survey Explorer ({\it WISE}), the Infrared Astronomical Satellite ({\it IRAS}), and the {\it Spitzer} instrument, the Multiband Imaging Photometer ({\it MIPS}) for NASA's Space Infrared Telescope Facility ({\it SIRTF}). Each instrument has differing operating capabilities, all of which are detailed below. 

{\it WISE} is an all sky survey mission covering the entirety of the northern and southern sky. The telescope obtains measurements in four bands (W1--W4) centered at 3.4, 4.6, 12, and 22\,$\mu$m respectively. The resolution in each of the bands is as follows: $6.1''$, $6.4''$, $6.5''$, \& $12''$ in each of the four bands \citep[W1, W2, W3, W4;][]{Wright10}.

{\it IRAS} surveys the sky at slightly longer wavelengths than {\it WISE}; its bands are centered at 12, 25, 60 and 100\,$\mu m$. We utilize the 12\,$\mu$m measurements for this work with a resolution of $0.5'$, notably larger than {\it WISE}. It is prudent to note that the {\it IRAS} 12\,$\mu m$ band has a width of 6.5\,$\mu m$, whereas the {\it WISE} 12 \,$\mu m$ has a band width of  5.5\,$\mu m$. This results in the {\it IRAS} measurements being on average slightly higher than the {\it WISE} measurements. 

{\it MIPS} images the sky with bands centered at: 24, 70, and 160\,$\mu$m. For this work we inspect the 24$\mu$m measurements with a given resolution of $6''$, comparable to {\it WISE}.

\subsection{Source Contamination}

Due to the angular resolution of {\it WISE}, and the significantly lower resolution of the {\it IRAS} instrument, source contamination due to galaxies in close proximity to our sources is possible. In order to diagnose any potential contamination, visual inspection of optical and infrared images of all sources in our sample was performed to assess if any contaminating objects lie within the field of view of the {\it WISE} or {\it IRAS} instruments. This inspection was paired with a literary search using the NASA/IPAC Extragalactic Database (NED) to identify the classification of any objects within $12''$ of our sources. For this search, we restrict our inspection to only galaxies and stellar objects as their infrared luminosities are significant enough to cause a detectable shift in the infrared measurements of our sources. We note that we do not consider any unresolved infrared sources as contaminating objects. 

In addition to the known interacting binary systems, $0941-080$ and $1934-638$, discussed in \citet{deVries98} and \citet{Jauncey86} respectively, we conclude that contamination in the {\it WISE} bands is possible in 8 of the 29 sources in our sample. Out of these, the following sources are expected to only have contamination in the longest-wavelength {\it WISE} band, W4 (resolution of $12''$), which we do not use for our color classification or in further analysis: $0019-000$, $1031+567$, $1117+146$ and $1323+321$. Therefore, for these four sources, the {\it WISE} color classification is expected to be unaffected by the nearby galaxies/stars, however, there is a potential for mild contamination in their {\it IRAS} fluxes. We note that from these only $1031+567$ has available {\it IRAS} flux measurements, and the difference in the measured flux between the {\it IRAS} and {\it WISE} is $~12\%$, see Table \ref{table:1}.

\begin{figure}[t!]
\centering
\includegraphics[width=\columnwidth]{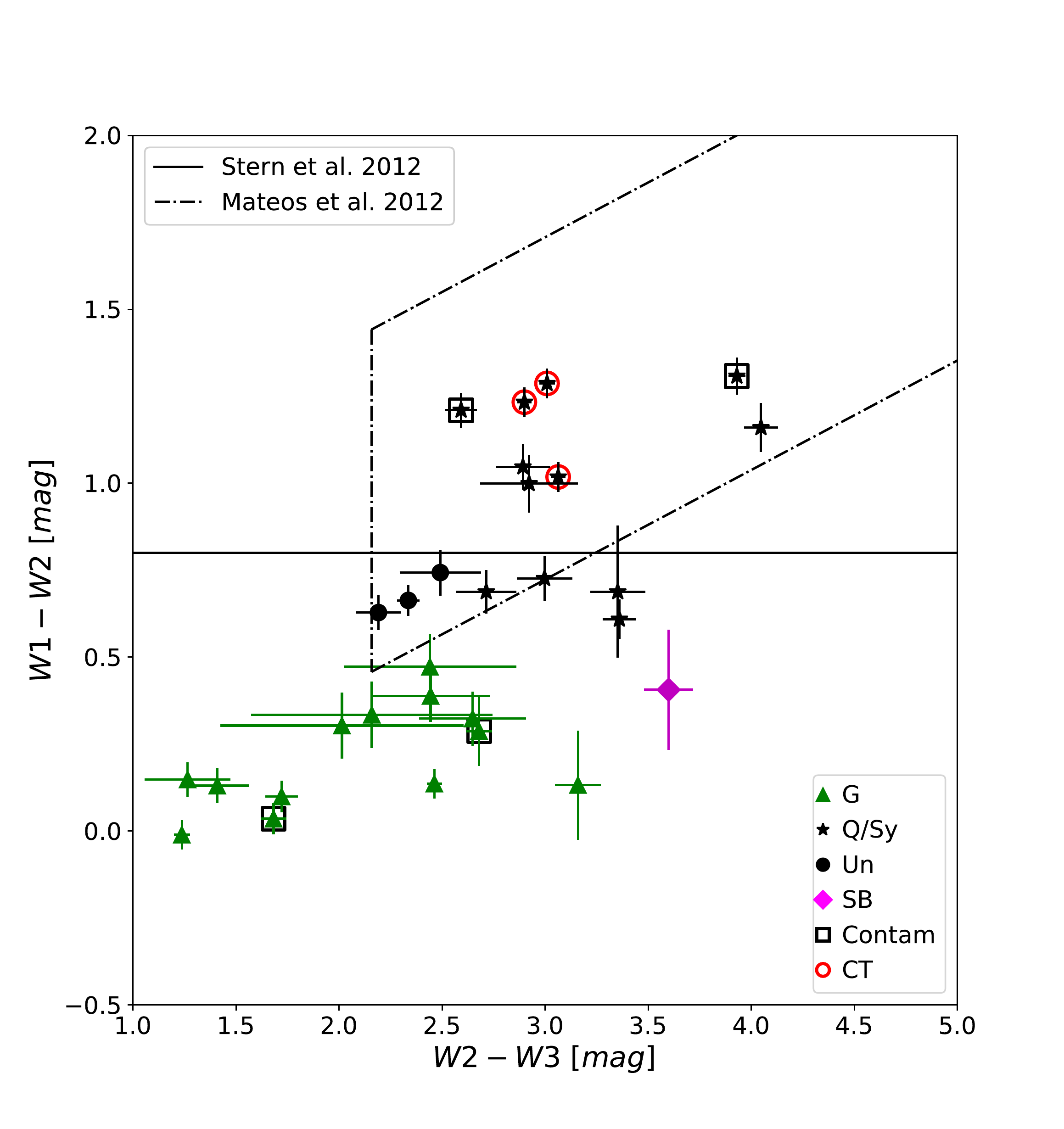}
\caption{{\it WISE} color-color diagram for compact radio galaxies included in our sample. Different symbols denote {\it WISE} color classification following \citet{Wright10}: ``Galaxy'' (G) -- green triangle, ``Starburst'' (SB) -- purple side-pointing triangle, ``Quasar/Seyfert'' (Q/Sy) -- black star, ``Uncertain'' (Un) -- black circle; sources marked with a black square may have a significant contamination from nearby background/foreground objects; red circles indicate objects confirmed as Compton-thick based on the X-ray spectroscopy. The horizontal solid line delineates the \citet{Stern12} cut: the MIR emission of sources located above the line is dominated by an AGN component. The dotted, dashed lines show the \citet{Mateos12} cut: the MIR emission of sources located within wedge is dominated by an AGN component.}
\label{fig:WISE}
\end{figure}

\begin{figure*}[t!]
\centering
\begin{tabular}{cccc}
\centering
	\includegraphics[width=4cm]{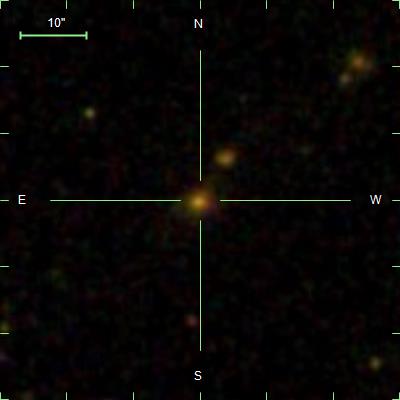} &
	\includegraphics[width=4cm]{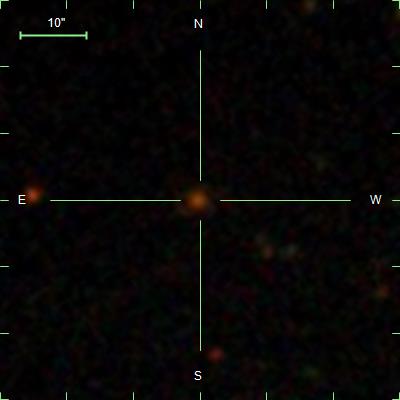} &
	\includegraphics[width=4cm]{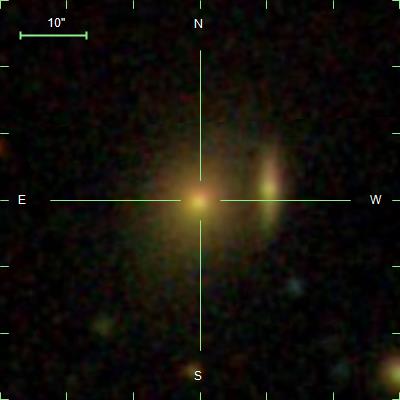} &
	\includegraphics[width=4cm]{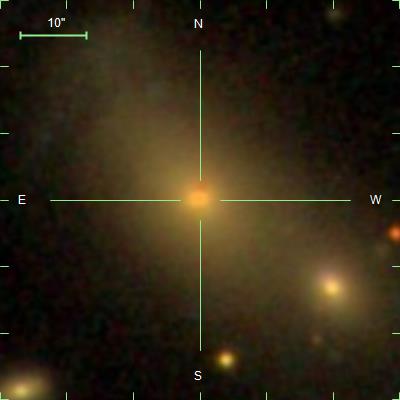} \\
    \includegraphics[width=4cm]{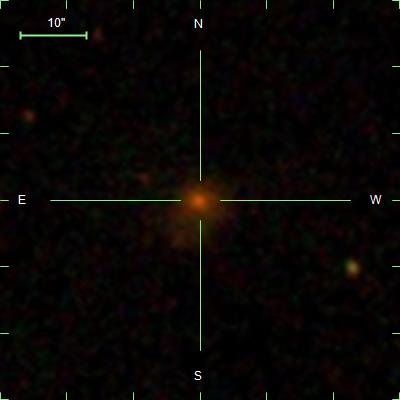} &
	\includegraphics[width=4cm]{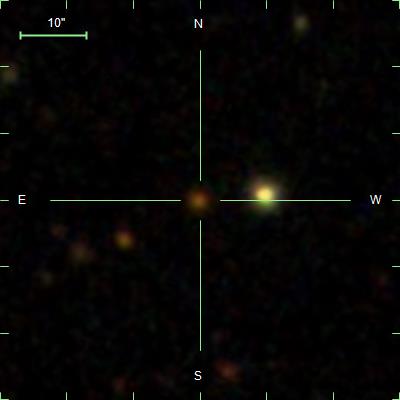} &
	\includegraphics[width=4cm]{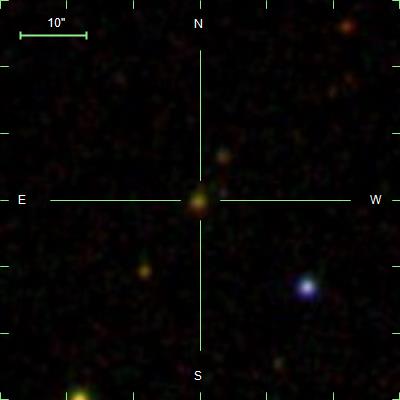} &
	\includegraphics[width=4cm]{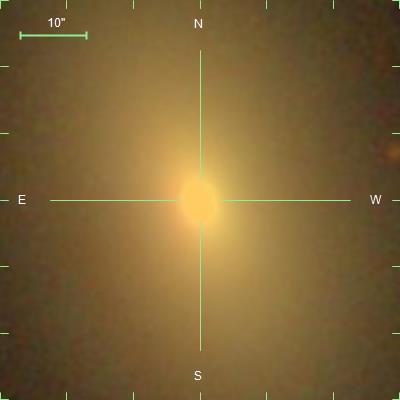} \\
   \includegraphics[width=4cm]{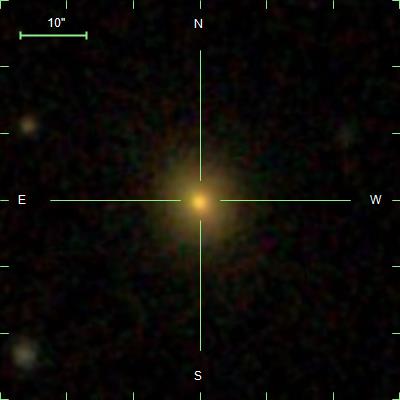} &
	\includegraphics[width=4cm]{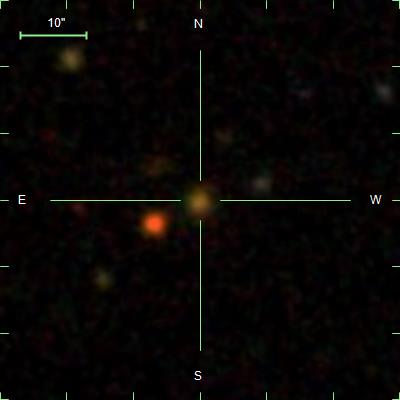} &
	\includegraphics[width=4cm]{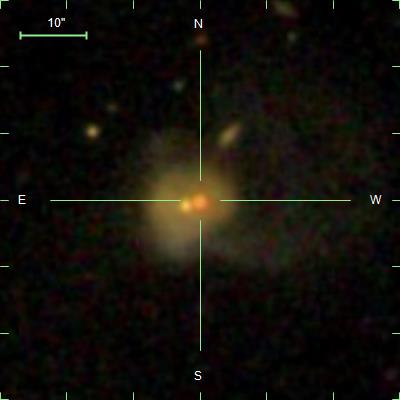} &
	\includegraphics[width=4cm]{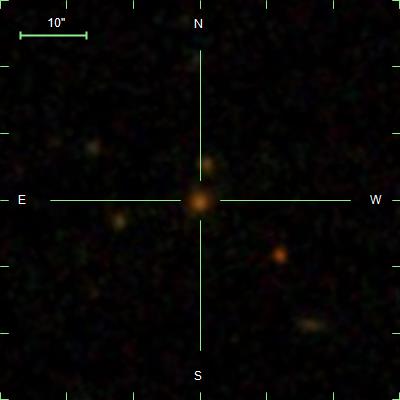} \\
    \includegraphics[width=4cm]{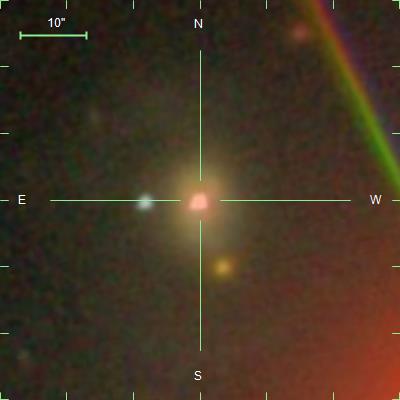} &
	\includegraphics[width=4cm]{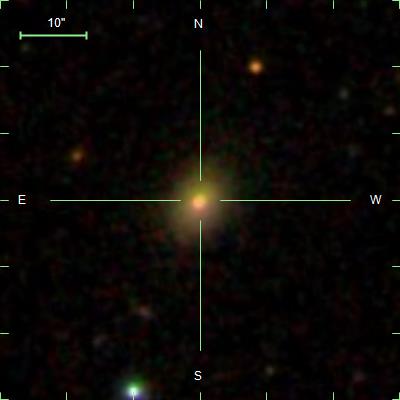} &
	\includegraphics[width=4cm]{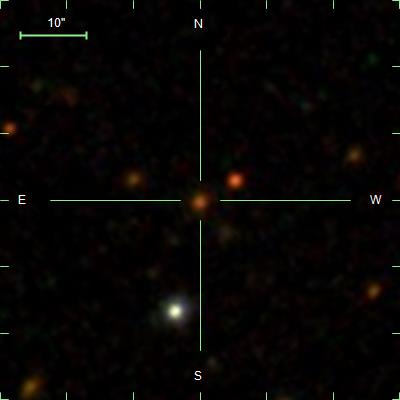} &
	\includegraphics[width=4cm]{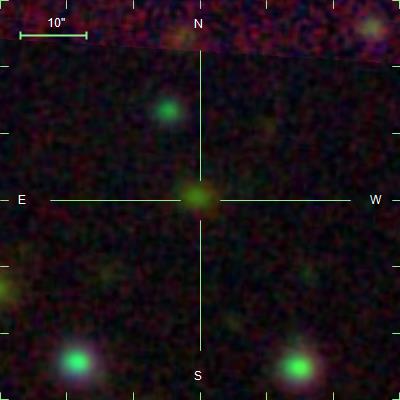} \\
\end{tabular}
	\caption{The SDSS gri color composite images of the host galaxies, available for 17 sources from our list \citep{Abolfathi18}, excluding the particularly faint 2128+048; these are (from left to right) 0019--000, 0026+346, 0035+227, 0116+319 (first row), 0428+205, 1031+567, 1117+146, 1146+596 (second row), 1245+676, 1323+321, 1345+125, 1358+624 (third row), 1404+28, 1511+0518, 1607+268, 2352+495 (fourth row). In each image, the position of the source is indicated by a cross, while the image scale is shown in the upper left corner, with each image being a square $60''$.}
\label{fig:sdss}
\end{figure*}

\begin{figure*}[t!]
\centering
\begin{tabular}{cccc}
\centering
	\includegraphics[width=0.24\textwidth]{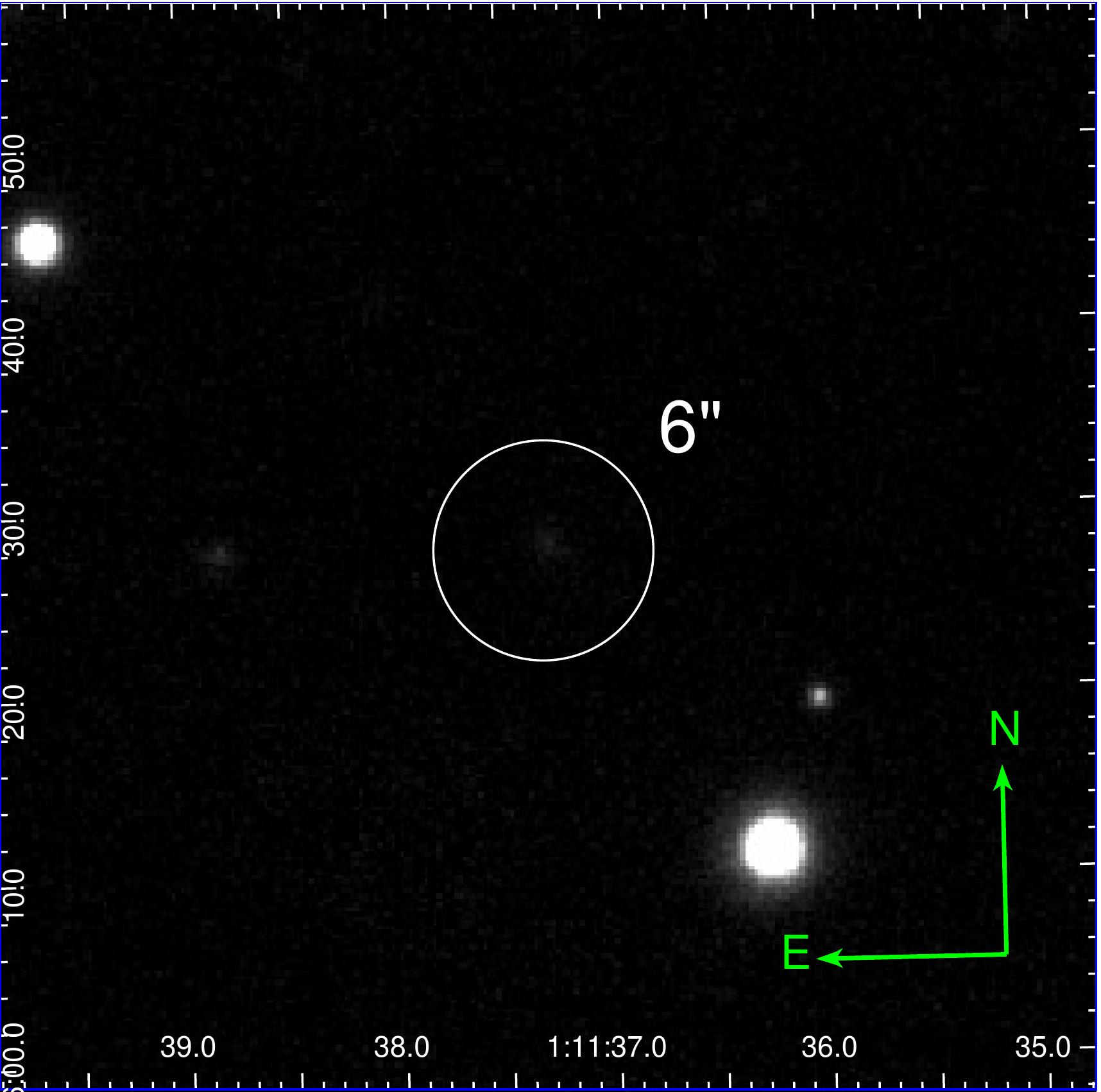} &
	\includegraphics[width=0.24\textwidth]{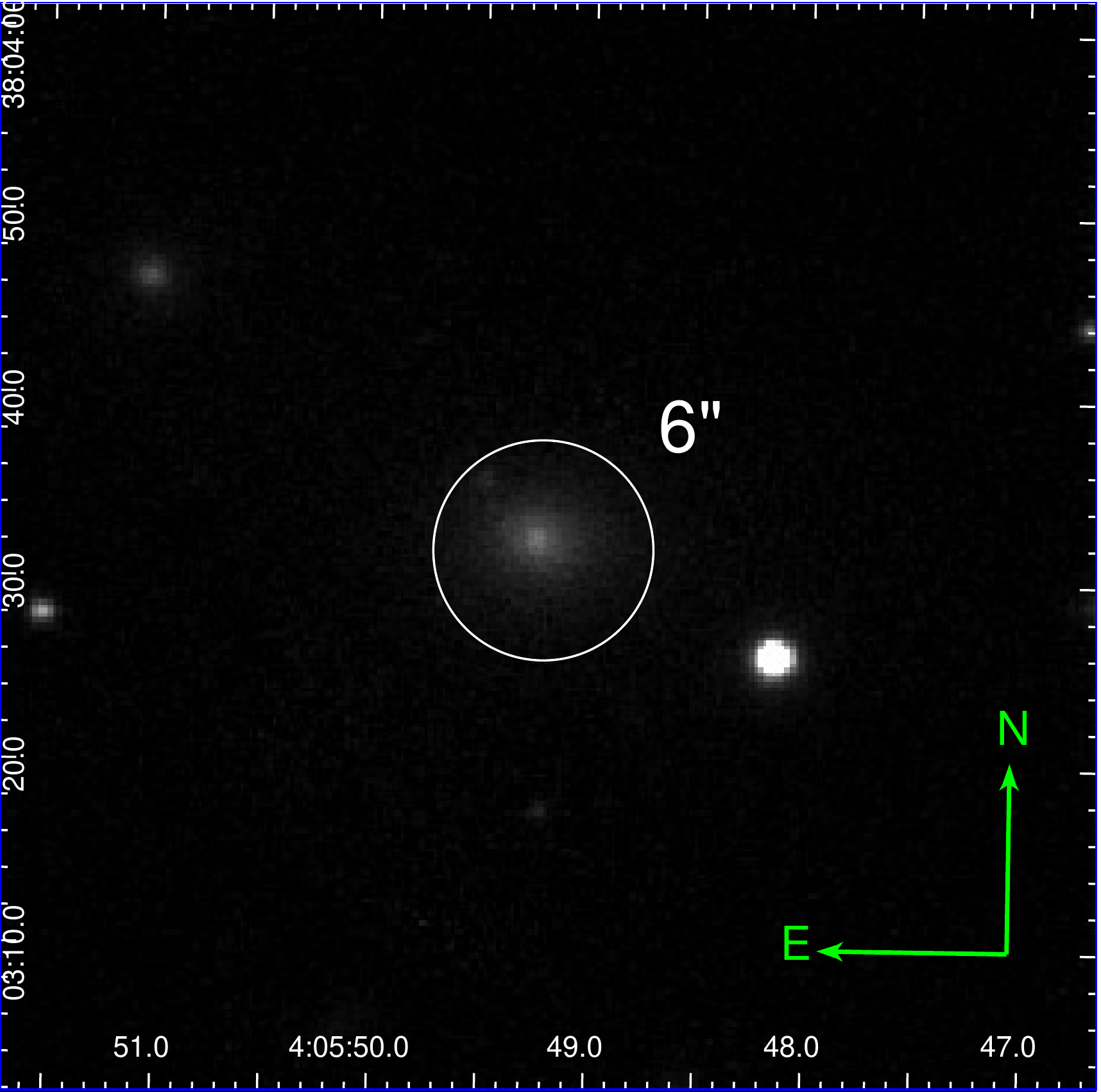} &
	\includegraphics[width=0.24\textwidth]{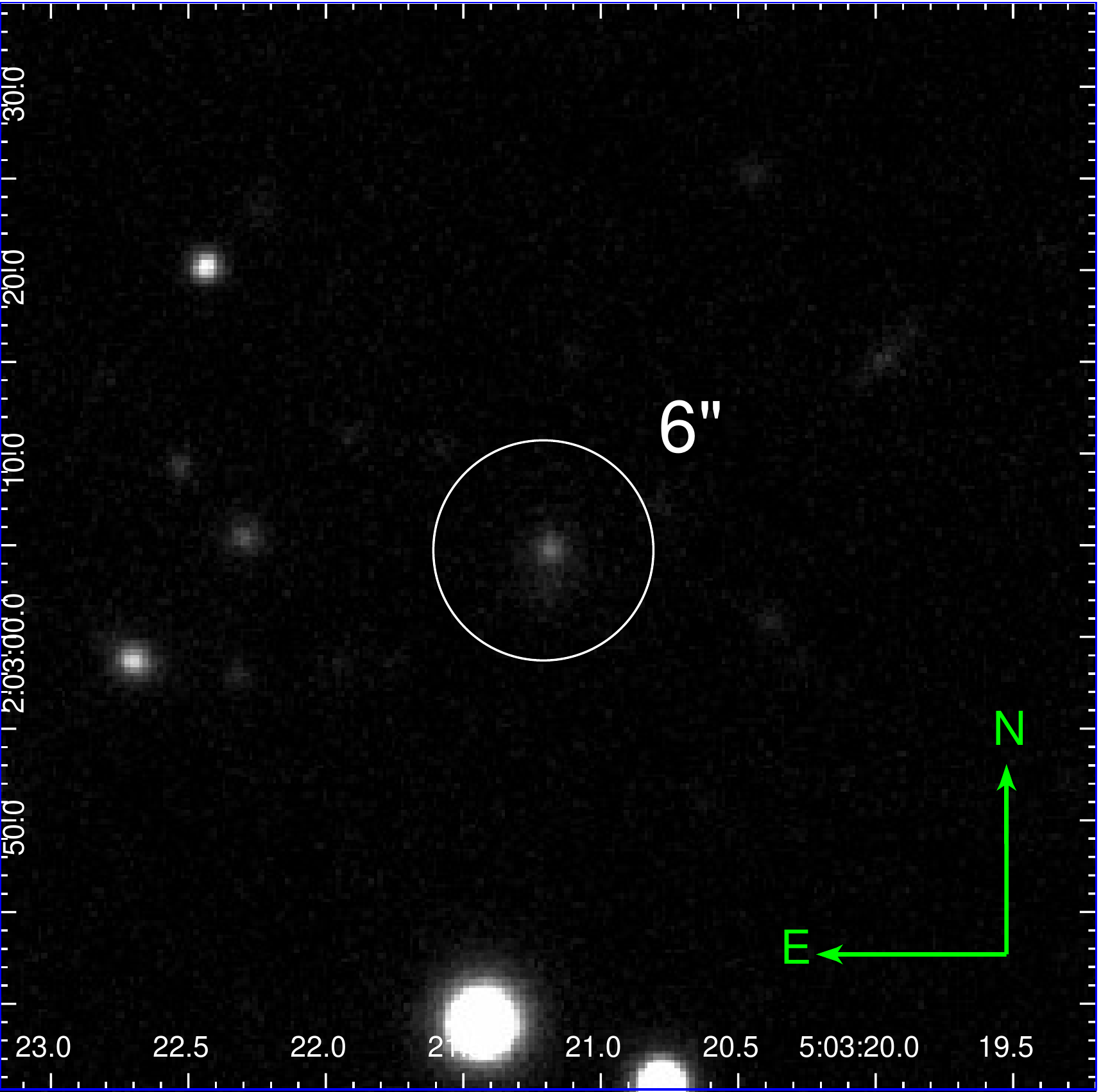} &
	\includegraphics[width=0.24\textwidth]{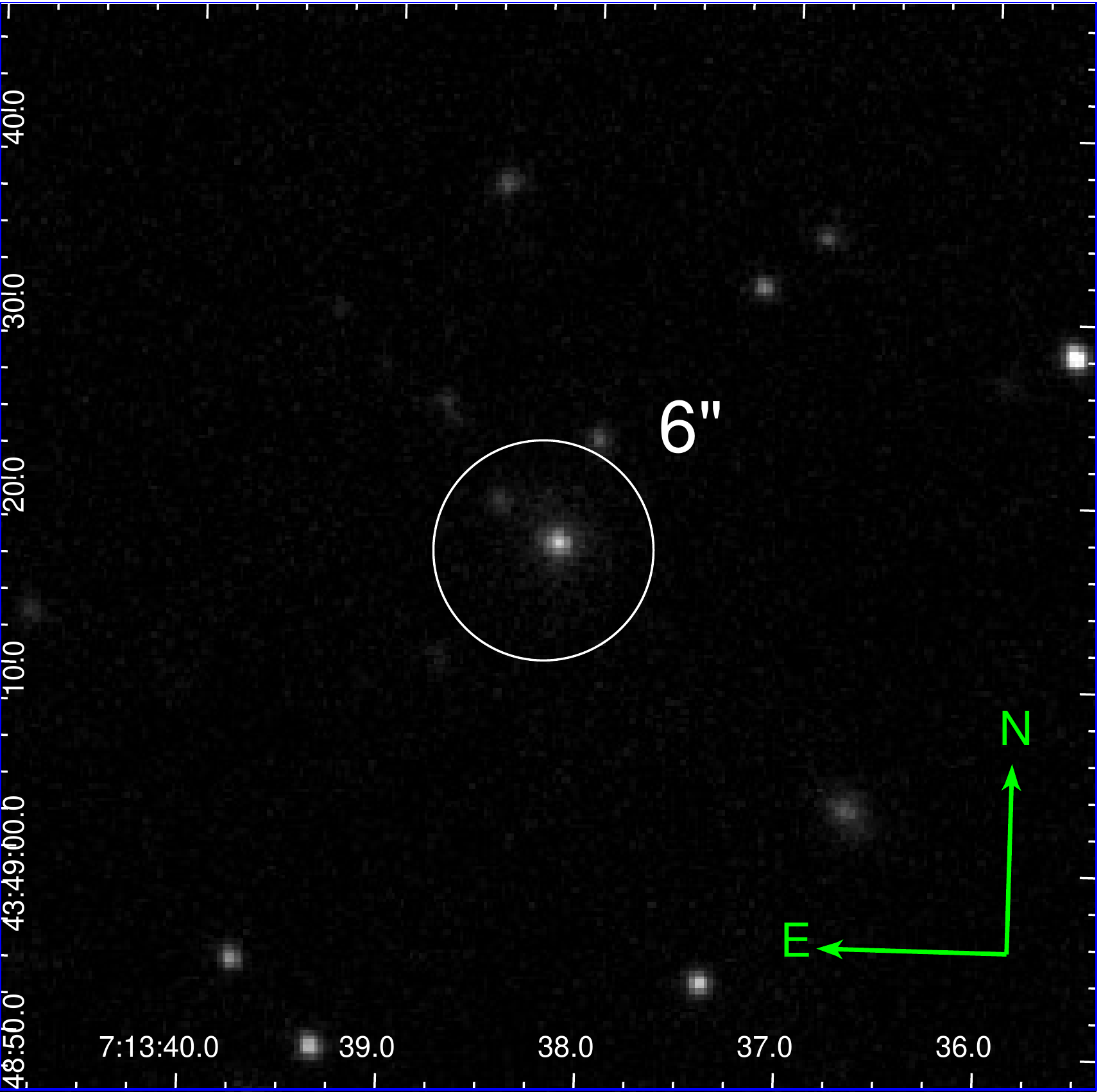} \\
	\includegraphics[width=0.24\textwidth]{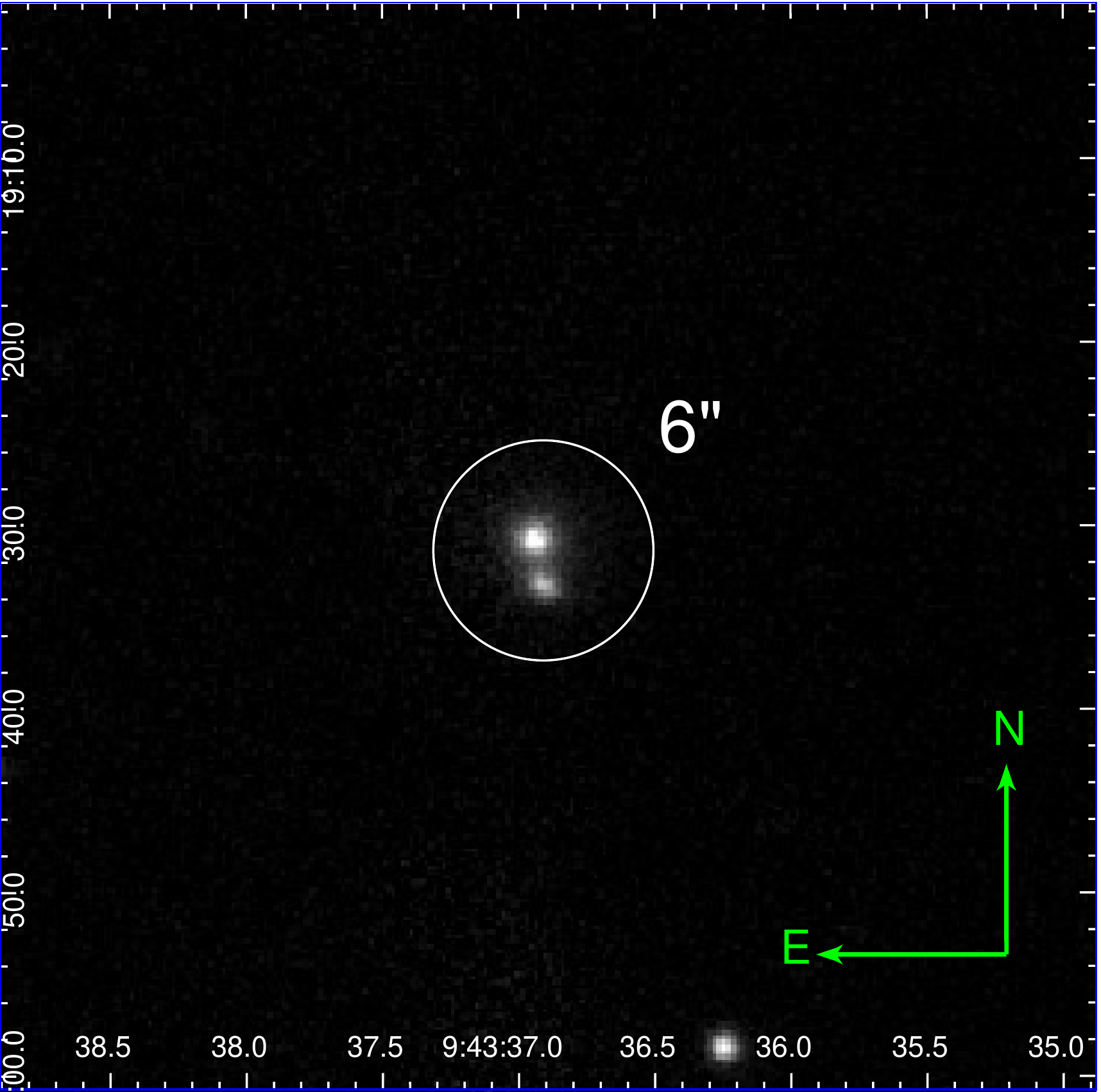} &
	\includegraphics[width=0.24\textwidth]{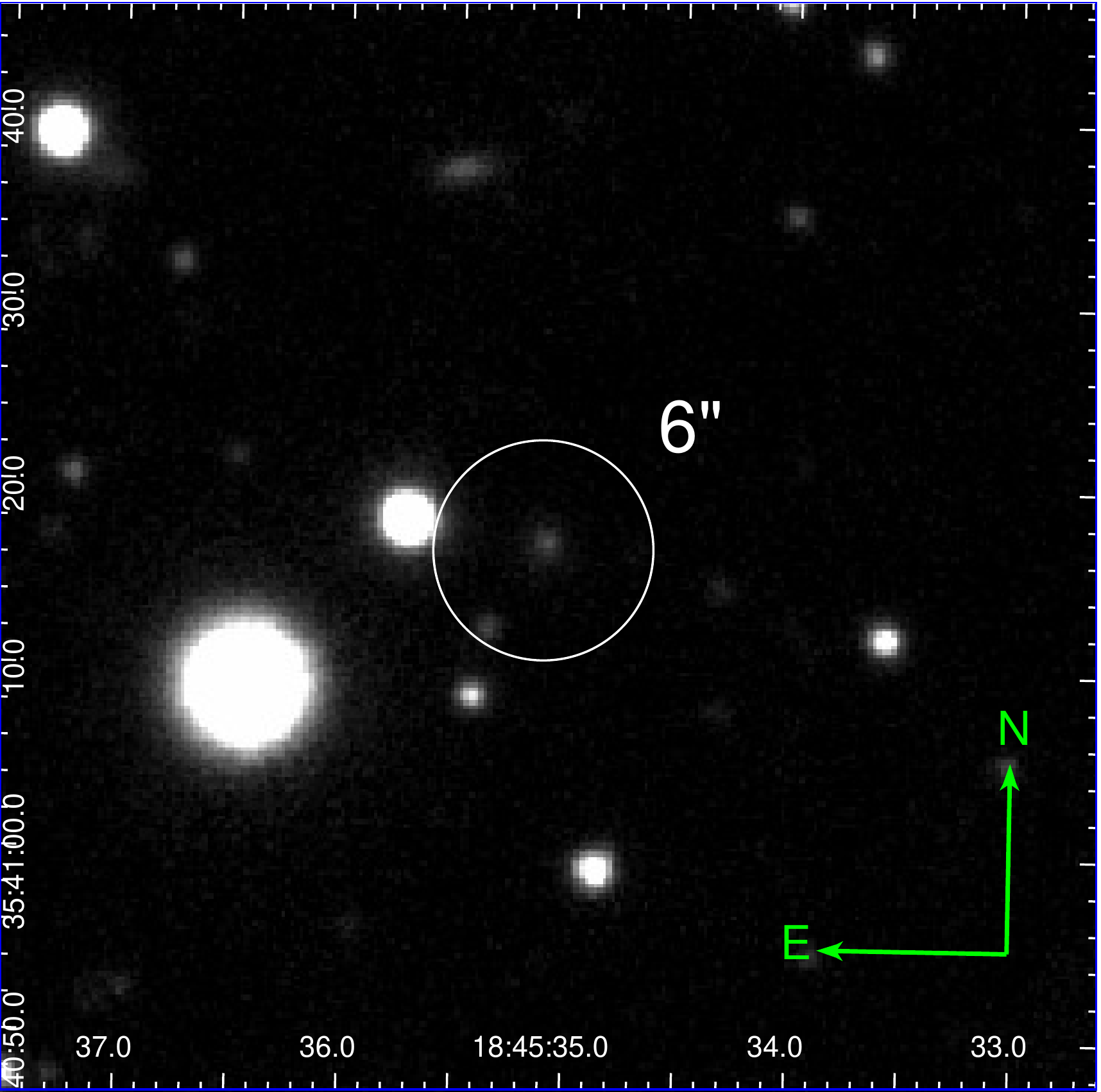} &
	\includegraphics[width=0.24\textwidth]{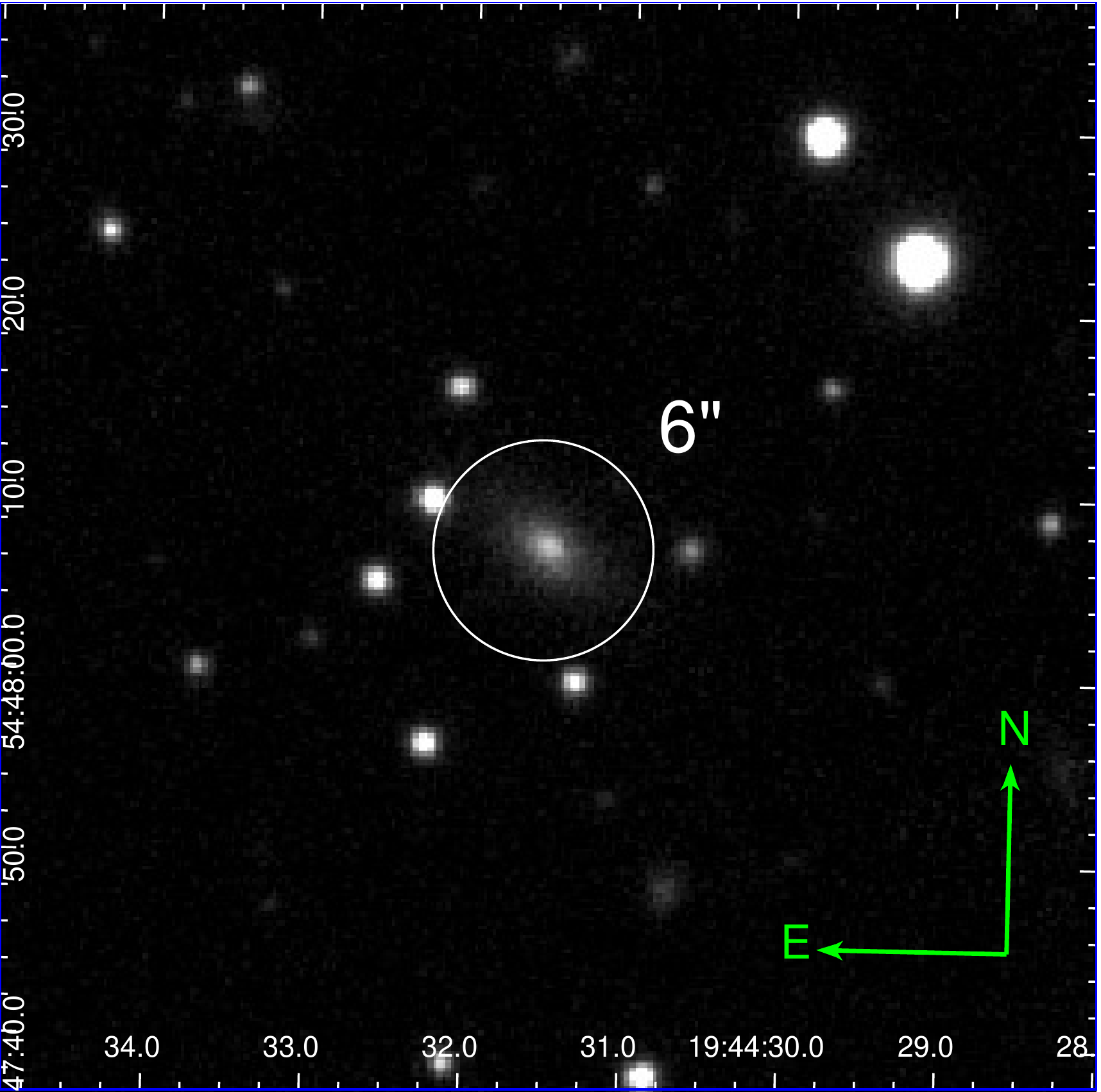} &
	\includegraphics[width=0.24\textwidth]{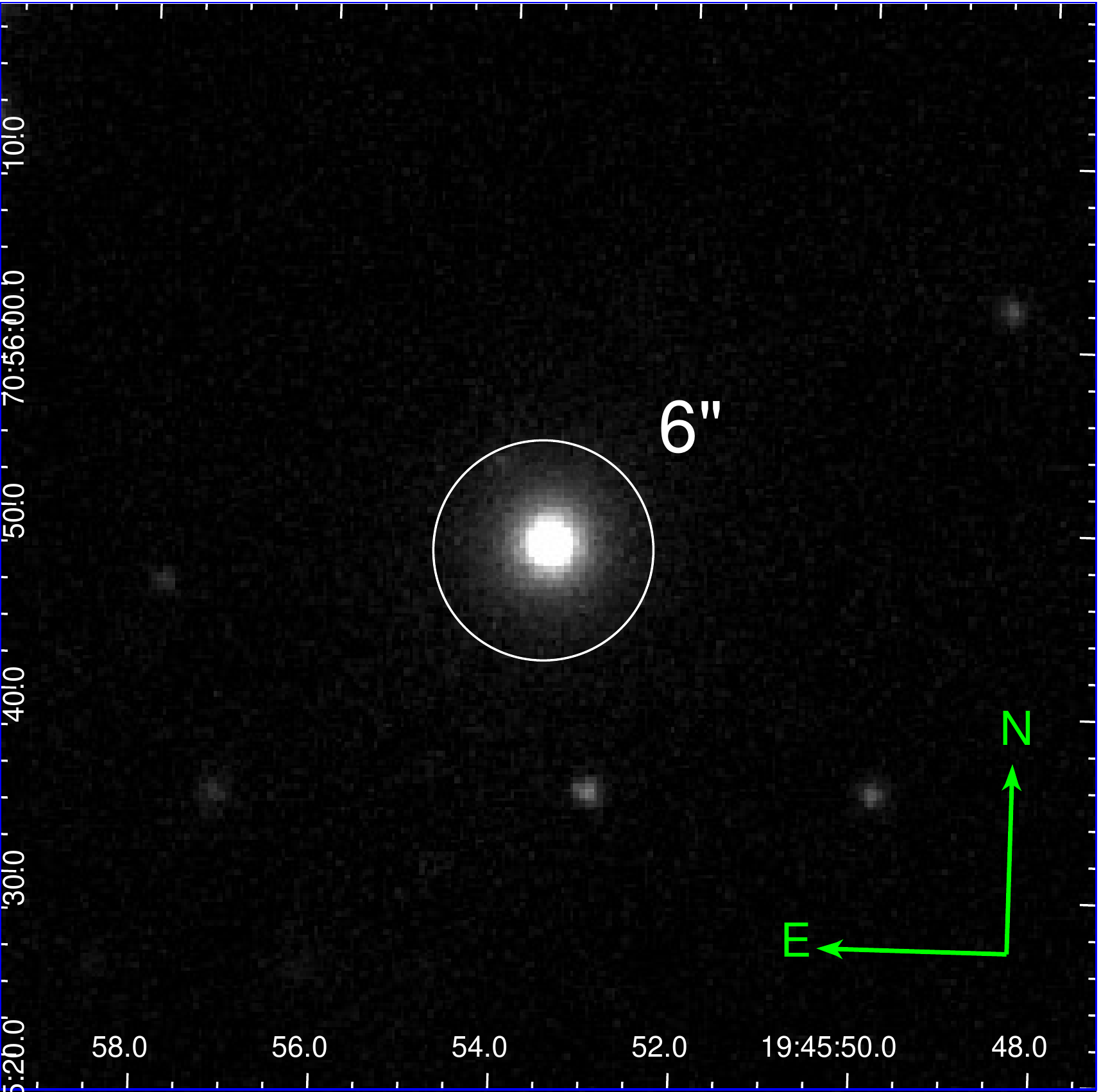} \\
	\includegraphics[width=0.24\textwidth]{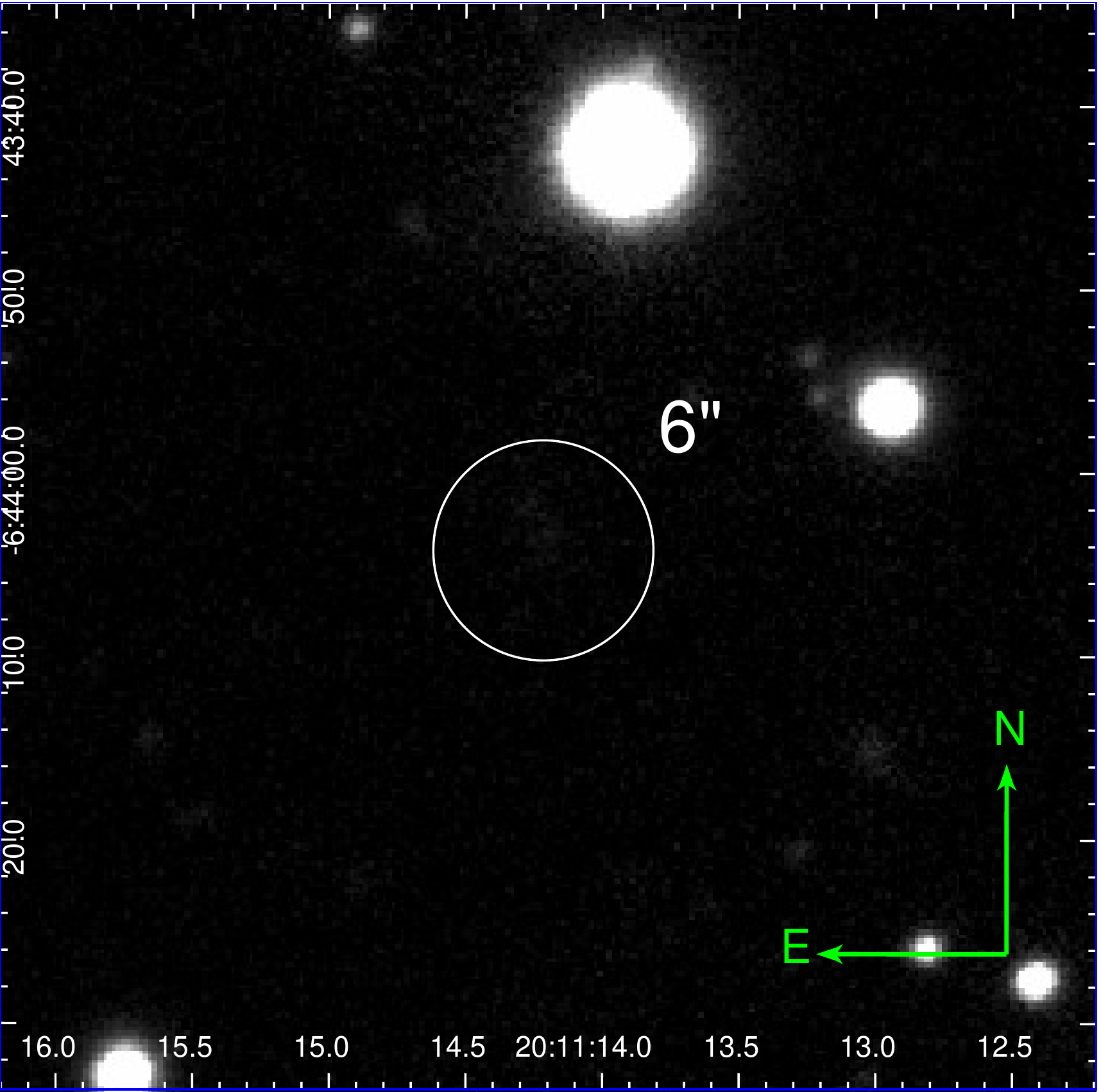} &
	\includegraphics[width=0.24\textwidth]{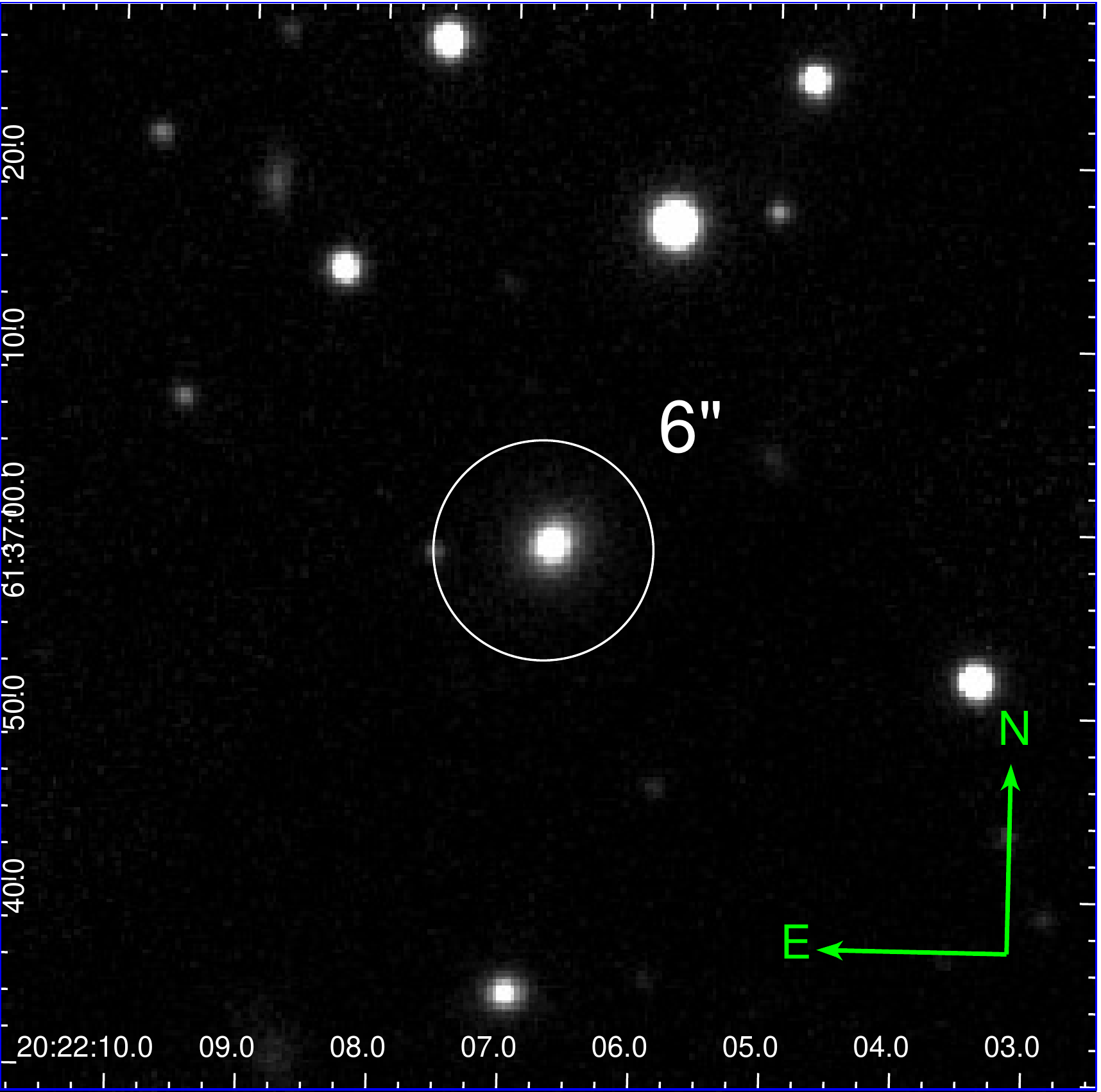} &
	\includegraphics[width=0.24\textwidth]{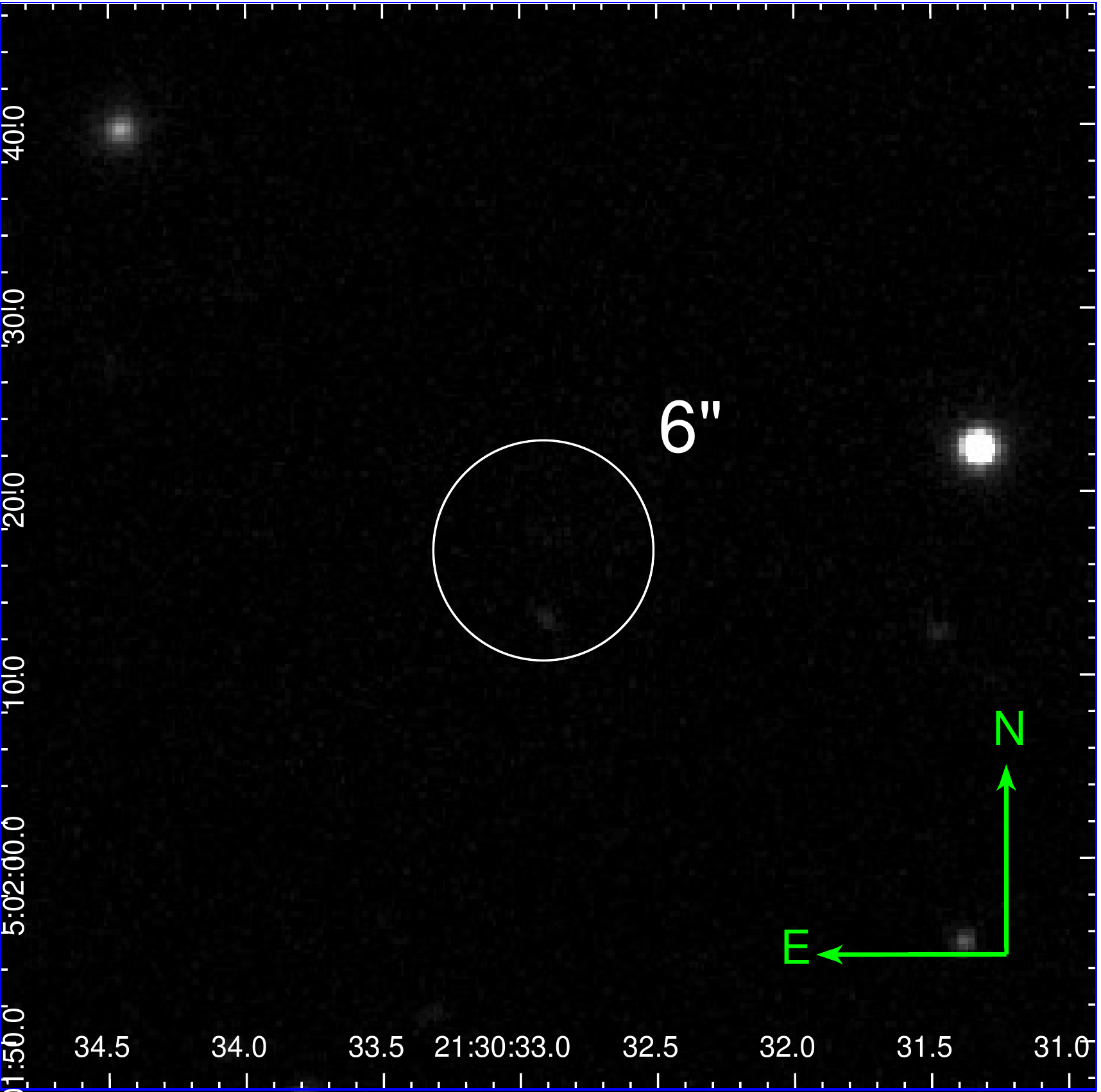} &
\end{tabular}
\caption{Optical images of sources not covered by SDSS (or too faint for SDSS), but observed instead by the all sky survey PanSTARRS, in the i band (7563\,{\AA}), including (from left to right) 0108+388, 0402+379, 0500+019, 0710+439 (first row),  0941--080, 1843+356, 1943+546, 1946+708 (second row), 2008--068, 2021+614, and 2128+048 (third row). In each image with overlaid ecliptic coordinate grids, the position of the source is indicated by a circle with a diameter of $6''$ with each image being a square $60''$.}
\label{fig:Pan}
\end{figure*}

\begin{figure}[b!]
\centering
\includegraphics[width=4.2cm]{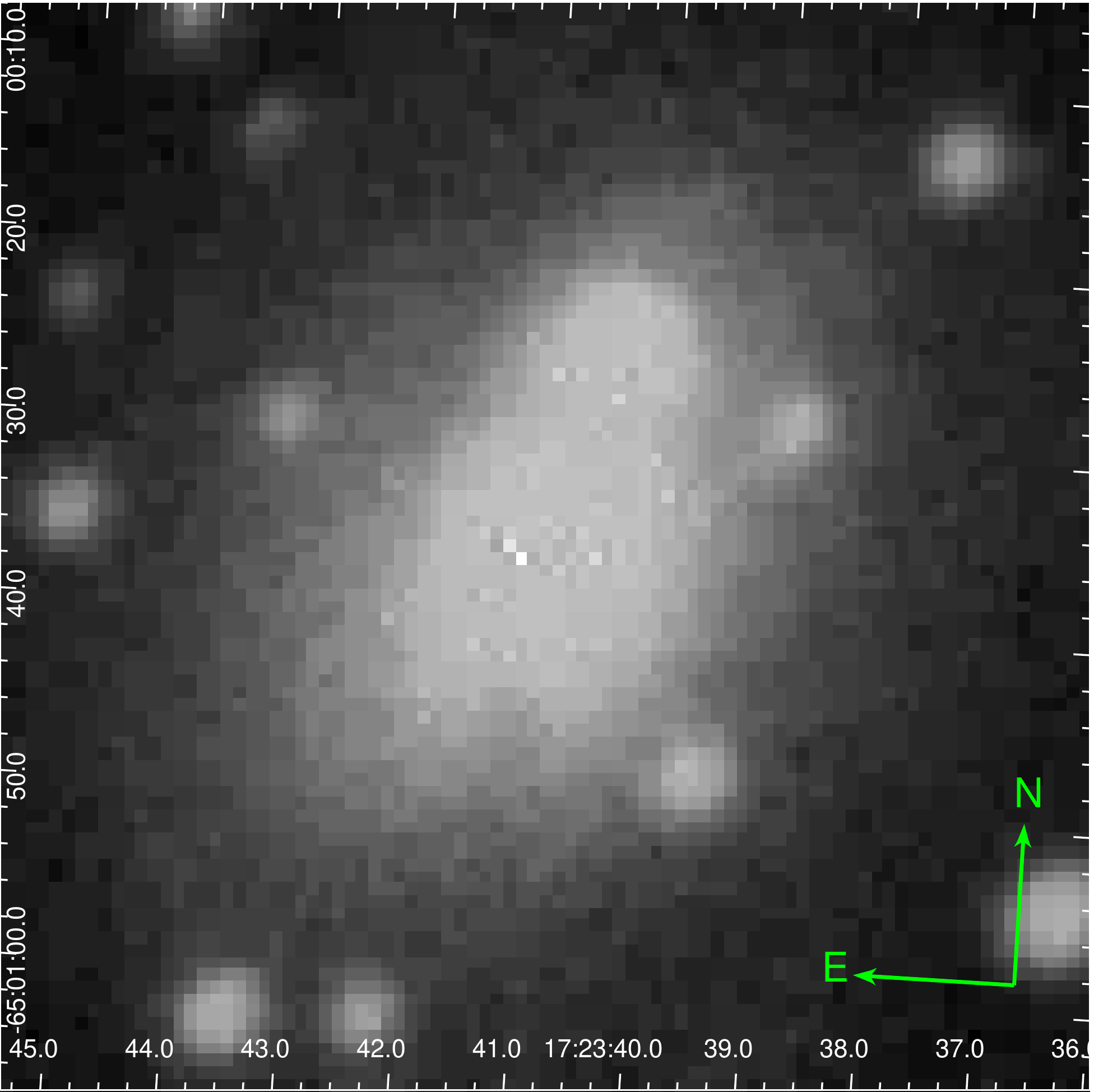}
\includegraphics[width=4.2cm]{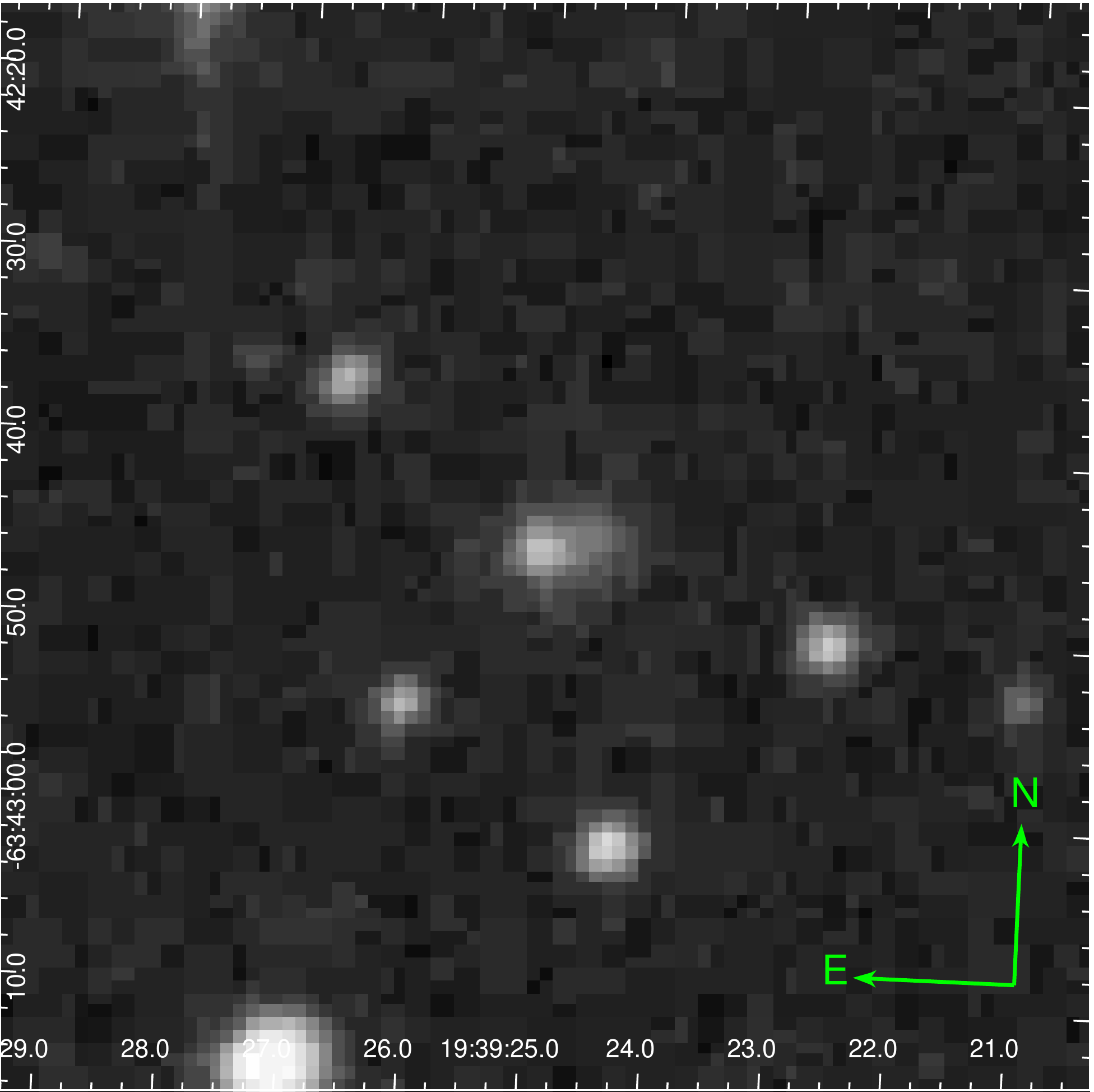} 
\caption{Additional optical images in the UKST blue band of the two southernmost sources 1718--649 (left panel) and 1934--638 (right panel), obtained from the Super Cosmos Sky Surveys. In each image, the position of the source is in the center of the image with each image being a square $60''$ and containing overlaid ecliptic coordinate grids.}
\label{fig:Cosmos}
\end{figure}

Contamination in the {\it WISE} color classification for the following sources is likely: $0116+319$, $1345+125$, $1358+624$, and $1607+268$, all marked in Table\, \ref{table:1}. These sources have galaxies and/or stars closer than $6.5''$, and as such would contain some level of excess flux in all four {\it WISE} bands and result in a shift in the positioning on our diagnostic plot. These four cases can be viewed in Figure\,\ref{fig:contamination}, which displays the {\it WISE} 12\,$\mu$m image and the respective Sloan Digital Sky Survey (SDSS) color composite image of each source. Optical images are color composite images of the g(4770\,{\AA}), r(6231\,{\AA}), and i(7625\,{\AA}) bands obtained through the SDSS sky server.  It can be clearly seen that the {\it WISE} images show a blending or merging of nearby objects that in the SDSS image are more clearly seen as separate objects. For the specific case of $1607+268$ which has a very low flux in the 12\,$\mu$m band; we provide the $3.4\mu$\,m band image as well. The morphology of  $1345+125$ is discussed in detail in \citet{Axon00}.

We finally note one source in particular, $2021+614$, which shows a significant difference between the {\it IRAS} and {\it WISE} flux measurements. The calculated {\it IRAS} luminosity is $~223\%$ higher than the respective {\it WISE} luminosity. This is most likely due to unresolved infrared sources near to $2021+614$ as well as poor data quality provided by {\it IRAS}. We comment, however, that the difference in the fluxes provides little change to our conclusions. 

\subsection{MIR Photometry}

The {\it WISE} colors, W1--W2 and W2--W3, can be utilized to diagnose the suggested level of star formation within a source and suggest the type of source being observed. The {\it WISE} colors for all sources in our sample have been estimated in a standard manner based on photometry in the W1 (3.4\,$\mu$m), W2 (4.6\,$\mu$m), and W3 (12\,$\mu$m) bands. The W1--W2 and W2--W3 color values are provided in Table\,\ref{table:1}, and plotted in Figure\,\ref{fig:WISE}. Based on those colors, and following the diagnostic plot provided in \citet{Wright10}, we have grouped our sources into distinct types denoted as ``Galaxy'' (G), ``Starburst'' (SB), Quasar/Seyfert (Q/Sy), or --- if the color falls in a non-defined region of the diagnostic plot -- ``Uncertain'' (Un). G type sources encompass the region of the plot associated with star formation levels typical of Elliptical and Spiral morphology. 

Infrared magnitudes were obtained for all 29 sources from {\it WISE}, and their luminosities were calculated using standard flux relations. In addition, 12\,$\mu$m were obtained from {\it IRAS} measurements for 16 out of the 29 sources, with the corresponding luminosities provided in Table\, \ref{table:1}. We provide the 12\,$\mu$m {\it IRAS} and 24\,$\mu$m {\it MIPS} luminosities for comparison with the obtained {\it WISE} values. We note that all but four sources have comparable 24\,$\mu$m {\it MIPS} luminosities to the 12\,$\mu$m {\it WISE} and {\it IRAS} measurements. 0710+439, 1345+125, 1607+268, and 1934-638 show an increase of $140-210\%$ in the 24\,$\mu$m band, are all classified as Sy, and are all morphologically classified as CSO. We note these four sources for further investigation in a follow-up paper, as potentially interesting sources to inspect spectroscopically. 

\section{Results \& Discussion} 
\label{sec:results}

\begin{figure*}[th!]
\centering
\includegraphics[width=0.45\textwidth]{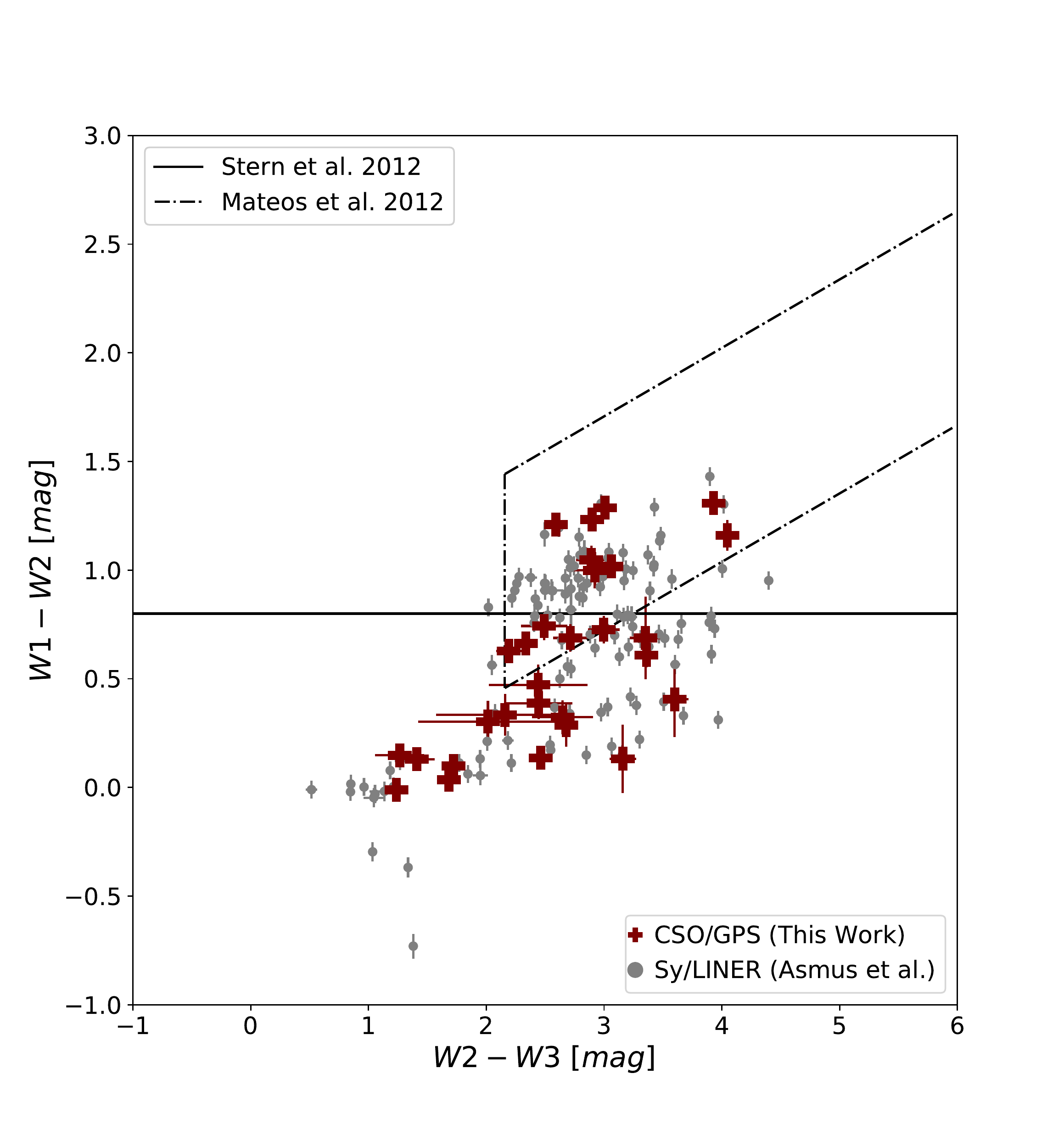}
\includegraphics[width=0.45\textwidth]{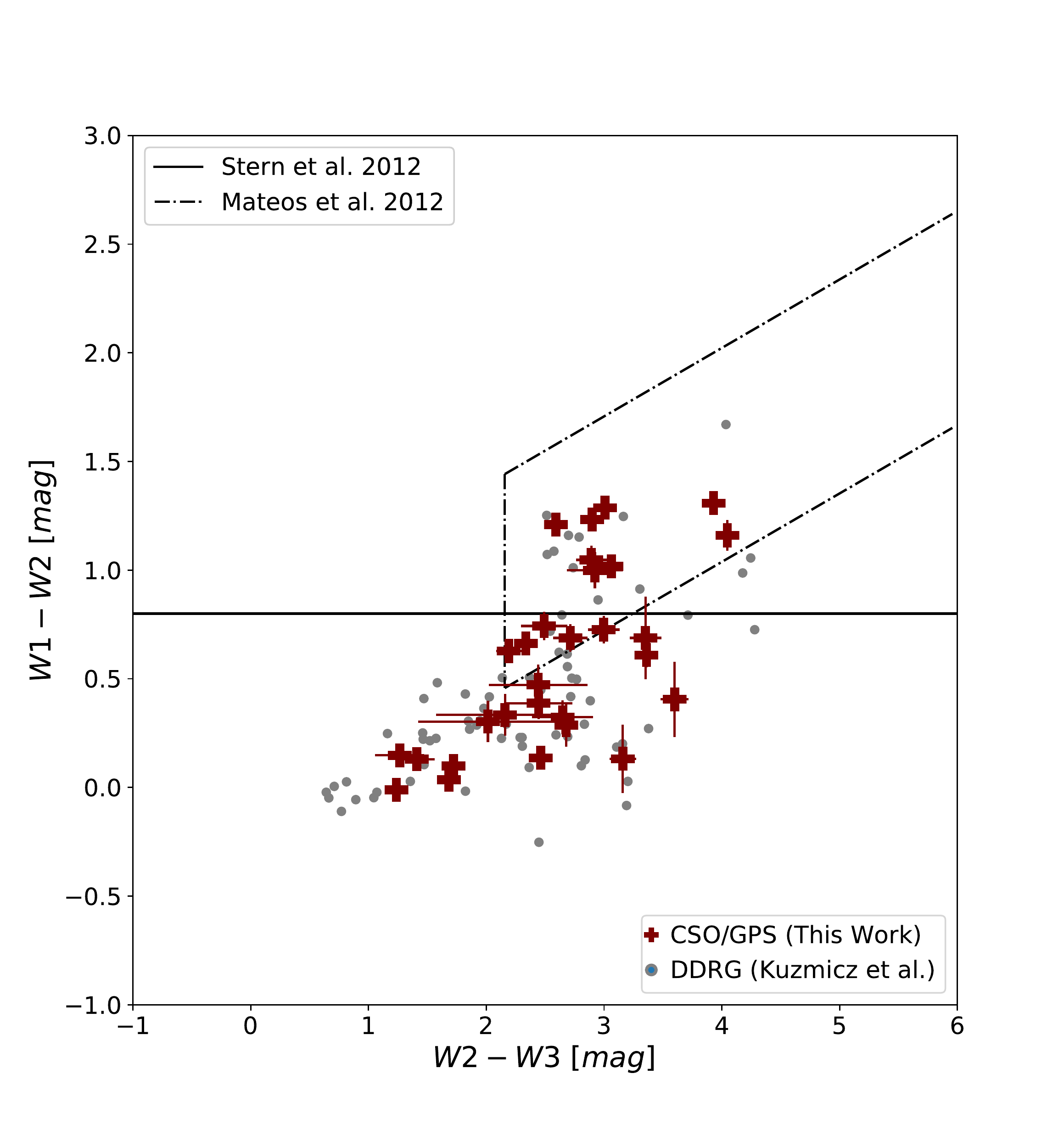}\\
\includegraphics[width=0.45\textwidth]{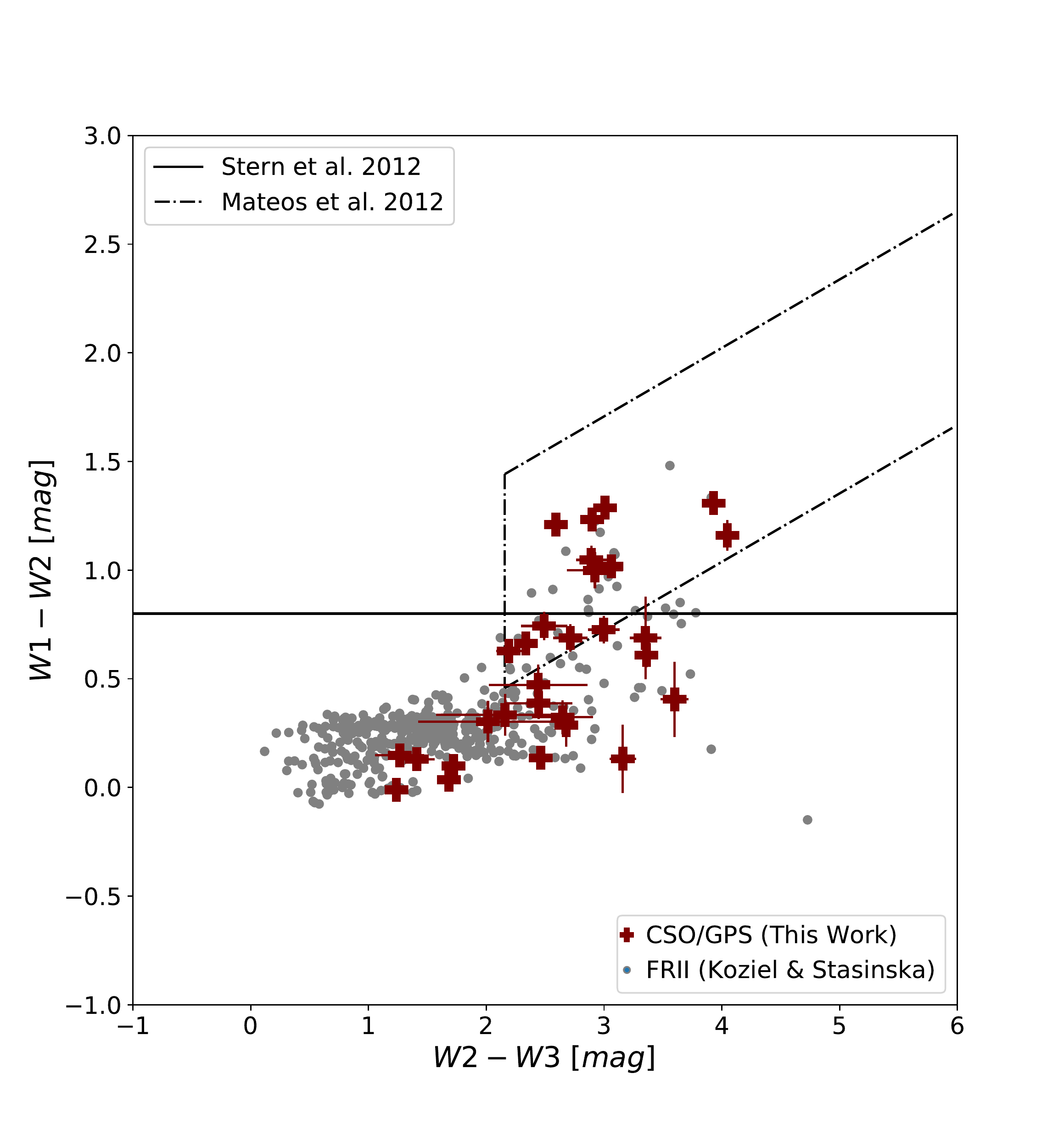}
\includegraphics[width=0.45\textwidth]{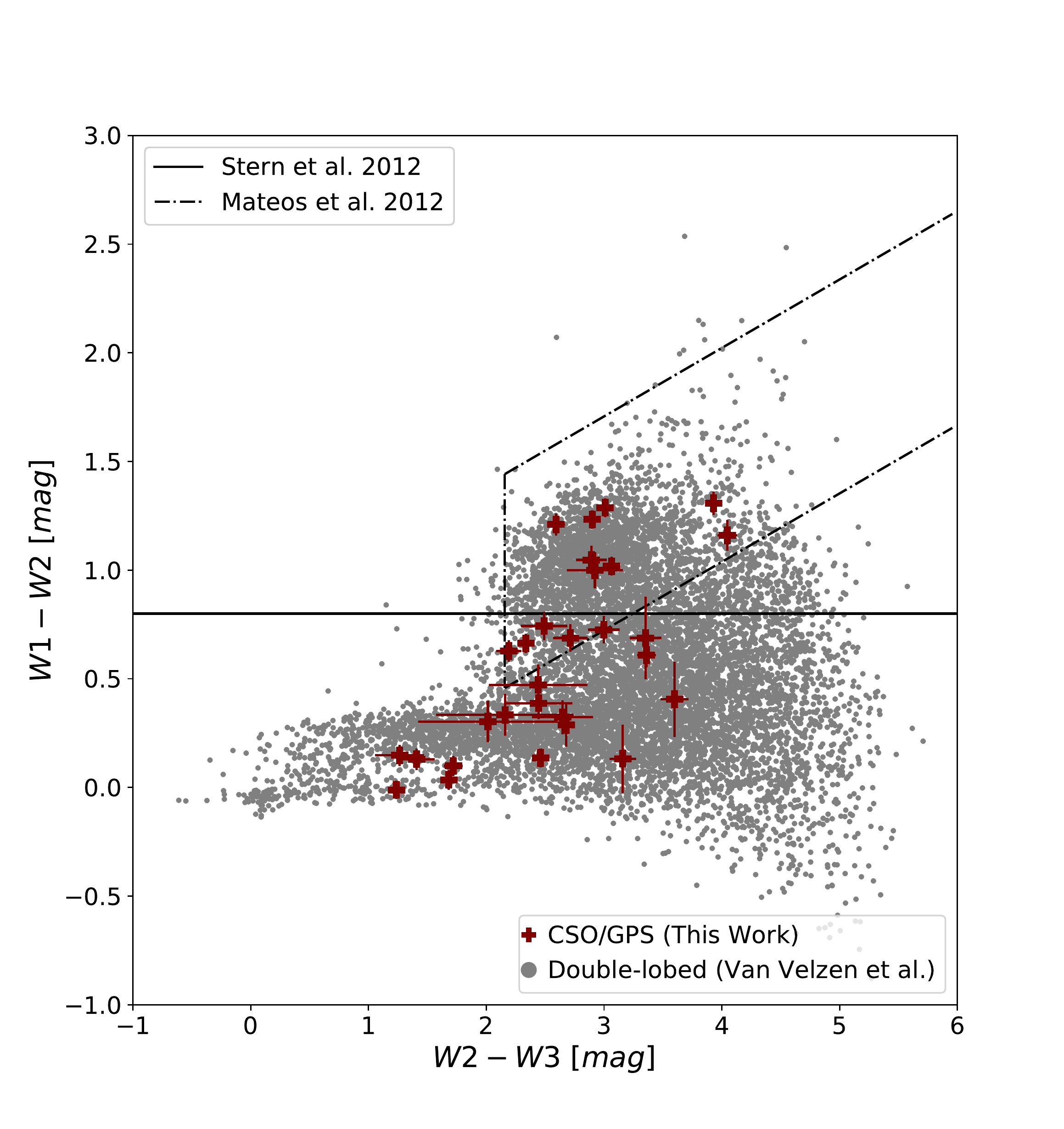}\\
\caption{ Top left panel: A comparison of the \citet{Asmus14} AGN sample of local Seyferts and LINERs with our sample of 29 GPS/CSO sources. Top right panel: A comparison of the \citet{Kuzmicz17} sample of restarted radio galaxies (DDRGs) with our sample of 29 GPS/CSO sources. Bottom left panel: A comparison of the \citet{Koziel11} sample of evolved classical doubles (FR\,II-type radio galaxies), with our sample of 29 GPS/CSO sources. Bottom panel: A comparison of the \citet{VanVelzen15} double-lobed radio sources sample with our sample of 29 GPS/CSO sources. The displayed cuts in all panels are the same as in Figure \ref{fig:WISE}. Larger red crosses in each panel indicate the objects in our sample with small grey circles in each plot denoting the respective comparison sample.}
\label{fig:WISE2}
\end{figure*}

\subsection{{\it WISE} Colors and Host Galaxy Morphologies}
\label{sec:WISE}

Figure\,\ref{fig:WISE} presents the {\it WISE} color diagnostic plot, with different symbols reflecting the {\it WISE} color identification following \citet{Wright10}. Clearly, such an identification does not serve as a rigorous classification, as it only points out the most likely origin of the detected MIR fluxes, corresponding to either the ISM with various levels of the star formation activity (G, SB) or the AGN component (Q/Sy).\footnote{Note in this context the overlaps between different types of galaxies in the {\it WISE} color classification: for example 1345+125, denoted here as ``Sy'', can additionally be classified as Ultraluminous Infrared Galaxy \citep[ULIRG; e.g.,][]{Rodriguez09}.} Quasar and Seyfert like galaxies are collected into one group due to significant overlap of their individual regions in the color--color diagram, however, we differentiate these sources in Table \ref{table:1}. This identification is, on the other hand, consistent with the selection cut introduced by \citet{Stern12}, and represented in the figure by a solid horizontal line: all the young radio galaxies with the {\it WISE} colors of the Q/Sy type, are indeed located either very close to the line, or much above it, and this substantiates the dominant AGN (dusty torus) contribution to the MIR emission in their cases. The overwhelming majority of our sources with the {\it WISE} classifications of Q/Sy, as well as all classified as Uncertain, fall within the \citet{Mateos12} cut, which constitutes a reliable MIR color selection of luminous AGN, constructed based on the flux-limited wide-angle Bright Ultrahard XMM-Newton Survey (BUXS; $4.5-10$\,keV flux limit of $6 \times 10^{-14}$\,erg\,s$^{-1}$\,cm$^{-2}$). 

Let us comment here on a few cases from our sample with ``uncertain'' {\it WISE} color identification. These sources lie, in fact, within the so-called ``blazar strip'' \citep{Massaro12}, which may suggest that their MIR fluxes are dominated by a non-thermal jet-related emission. Unlike in blazar sources, however, such an emission does not have to be relativistically beamed, but instead may constitute a high-energy tail of the synchrotron continuum of compact radio lobes expanding with sub-relativistic ($\lesssim 0.1 c$) velocities \citep[see in this context the broad-band modelling by][]{Ostorero10}.

Interestingly enough, all the objects in the sample confirmed as Compton-thick, based on the X-ray spectroscopy \citep[1404+286, 1511+0518 and 2021+614; see][respectively]{Siemiginowska16,Sobolewska19a,Sobolewska19b}, are characterized by the {\it WISE} colors as being consistent with a quasar identification, as in fact expected. Hence, {\it WISE} color diagnostics could, in general, be considered as a very useful tool when selecting candidates for Compton-thick AGN among compact radio galaxies.

The most striking feature of the diagram is, however, the fact that the most compact radio galaxies with colors W1--W2\,$\leq 0.5$\,mag, i.e. those presumably dominated in MIR by the ISM radiative output, populate almost uniformly the entire region occupied by galaxies (from elliptical to starbursts) with the W2--W3 colors between 1.0 and 4.0, suggesting a wide range of pronounced star formation activity within their hosts. Such a diversity could imply that the triggering of radio jets in AGN is not differentiated between hosts with substantially different fractions of young stars.

In this context, we investigate the optical images of the host galaxies in our sample. SDSS images of the host galaxies \citep{Abolfathi18} are available for 17 sources from our list; these are shown in Figure\,\ref{fig:sdss} (excluding the particularly faint 2128+048). In each image, the position of the source is indicated by a cross, while the image scale is shown in the upper left corner. These images are combined color images of the u, g, r, i, and z optical bands, obtained through the SDSS DR15 Navigate Tool. The SDSS images of the majority of our galaxies allow for an approximate morphological classification (the only exceptions are 2128+048, which is too faint to be classified, and 2352+495, whose image shows false colors). 
Most of the sources are red/yellow galaxies with an elliptical shape. Five galaxies have distorted morphology or show the signs of galaxy interaction; these are 0035+227, 0116+319, 0428+205, 1345+125, and 1404+286. None of the galaxies classified as "galaxy" in the MIR color diagram shows the indication of a spiral structure. On this basis, we can conclude that the MIR colors differentiate galaxies with different levels of star formation, but they are loosely related to morphological types. 

For the remaining sources which are not covered by the SDSS, or which are too faint for the SDSS, we retrieved the optical images of the hosts from the all sky survey PanSTARRS, in the i band (7563\,{\AA}), and, in the case of the two southernmost objects 1718-649 and 1934-638, from the Super Cosmos Sky Surveys in the UKST blue band; these are shown in Figures\,\ref{fig:Pan} and \ref{fig:Cosmos}, respectively. Again, most of those galaxies whenever bright enough to be classified morphologically, display elliptical shapes, with the exception of the peculiar 1718--649 host, {\it ``having the appearance of a high luminosity elliptical with faint outer spiral structure''} as noted by \citet{Fanti00}. We comment more on this object in Section\,\ref{sec:gamma}. These images are shown as strictly optical intensity and as such  no comments on their colors are made. We note that 0108+388, 2008--068, and 2128+048 are too faint in optical for any classification.

It is interesting to compare the distribution of the {\it WISE} colors in our sample of young radio galaxies with comparison samples of various types of active galaxies. For this purpose, we consider first the list of local (redshifts $z<0.4$) AGN compiled by \citet{Asmus14}, which includes 102 sources selected from the nine months of observations with the Burst Alert Telescope (BAT) onboard the {\it Swift} satellite within the $14-195$\,keV band (constituting a flux limited sub-sample in the list), complemented by the AGN with the available high-resolution MIR imaging enabled by ground-based 8m class telescopes such as VLT/VISIR, Gemini/Michelle, Subaru/COMICS, or Gemini South/T-ReCS. All together there are 253 sources in the \citeauthor{Asmus14} sample, for which the optical spectroscopic classification includes Seyferts type I, Seyferts type II, LINERs, and AGN/starburst composites. Moreover, the AGN from this sample are hosted by various types of galaxies, including late type galaxies with bona-fide spiral structures, not only elliptical or disk systems. 

The distribution of the {\it WISE} colors for the \citet{Asmus14} sample, in a direct comparison with our sample of GPS/CSOs, is shown in the top left panel of Figure\,\ref{fig:WISE2}. In order to establish a secure sample of {\it WISE} sources, we performed two quality cuts to remove any uncertain measurements. First we removed all sources that had a null or $<1$ signal to noise ratios (snr) for bands W1, W2, $\&$ W3. Following this cut, we removed any sources with flagged contamination (marked by $cc\_flags$ not equal to 0). These cuts left us with a sample of 121 sources from the original 253 sources in the \citeauthor{Asmus14} sample. Both distributions appear very similar and we quantify this statement by performing the two-dimensional Kolmogorov-Smirnov (2D KS) test, that was developed through the efforts to generalize the classical one-dimensional KS test to two \citep{Peacock83} and higher dimensions \citep{Fasano87}. In particular, we follow the 2D KS algorithm described in \citet{Fasano87} and calculate --- under the null hypothesis that the two analyzed samples were drawn from the same distribution --- the two-tailed $p$-value (i.e., the probability of obtaining a value of the statistic D greater than the observed value, if the null-hypothesis were true). As a result, we obtain $p=0.0795$ and $D=0.282$; adopting a significance level of $\alpha=0.05$, this result implies that the null hypothesis should be accepted.

We note that the \citet{Asmus14} sample includes both radio-quiet and also radio-loud objects --- either low-power radio galaxies of the Fanaroff-Riley type I (FR\,Is), or high-power classical doubles, i.e. Fanaroff-Riley type II radio galaxies (FR\,IIs) --- however, the radio properties were not taken into account when selecting the sample. That is, neither a detection of a radio counterpart nor a particular threshold value for the radio loudness parameter constituted a selection criterion. It is therefore interesting that the distribution of the {\it WISE} colors for this ``hard X-ray selected/quasi MIR selected'' sample of local AGN with the median redshift $z=0.016$, is statistically similar to the distribution of {\it WISE} colors in our sample of the youngest radio galaxies, selected based on their radio properties and availability of X-ray data, and in addition spanning a wider range of redshifts from $0.011$ up to $0.99$, with the median of $z=0.238$. We note, however, that we are looking at a smaller set of the \citet{Asmus14} sample, due to the quality cuts necessary for the {\it WISE} data, which could impose some unexpected selection effects.

Second, we compare the MIR colors of our GPS/CSOs with those characterizing the \emph{evolved} (extended) FR\,II type radio galaxies. In particular, we consider the FR\,II sample by \citet{Koziel11}, which resulted from cross-identification of the objects included in the Cambridge Catalogues of Radio Sources (3C--9C) with galaxies from the SDSS DR7 main galaxy sample, after excluding quasars and only keeping the objects with clear FR\,II large-scale radio morphology, investigated manually based on the available radio maps from the NRAO VLA Sky Survey (NVSS) and the Faint Images of the Radio Sky at Twenty-cm (FIRST) survey. This sample amounts to 401 sources spanning a redshift range from $0.045$ up to $0.6$, which is well matched to the redshift range in our GPS/CSO sample. 

The distribution of the {\it WISE} colors for the \citet{Koziel11} sample, in a direct comparison with our sample of GPS/CSOs, is presented in the top right panel of Figure\,\ref{fig:WISE2}. As shown, the analyzed GPS/CSOs, as a population, have significantly different MIR colors than FR\,IIs. Unlike compact radio galaxies, the evolved FR\,IIs typically cluster within the area of the color diagram occupied by elliptical galaxies with very little star formation activity. In addition, we note that the sample selection adopted by \citet{Koziel11}, specifically the removal of SDSS classified quasars, may result in a reduction of bright AGN in this sample. Again, we quantify our statement by running the 2D KS test, and obtain $D=0.6$ with the corresponding $p= 0.95\times 10^{-7}$, meaning that the null hypothesis stating that the two samples were drawn from the same distribution, can be safely rejected at the adopted significance level, $\alpha=0.5$. This result is in fact unexpected, as CSOs (at least the most luminous ones) are widely considered to be a young progenitor of FR\,II radio galaxies \citep[see, e.g.,][and references therein]{Perucho16}, and the evolution of radio structures from a compact ($<1$\,kpc) CSO phase to an evolved (linear sizes $\sim$\,10s--100s\,kpc) classical double phase, takes only up to $\sim 100$\,Myr, which is much shorter than the timescale required for host galaxy evolution from a late to an early type.

Third, we consider the sample of radio galaxies with confirmed recurrent jet activity, i.e. radio galaxies displaying clearly extended/outer radio lobes (predominantly of the FR\,II morphological type), in addition to a compact/inner double structure produced during the distinct, newborn jet phase. For these so-called ``double-double radio galaxies'' \citep[DDRGs; see][]{Schoenmakers00}, we use the list recently compiled by \citet{Kuzmicz17}, including 73 sources within a redshift range $0.002 < z < 0.7$. The distribution of the {\it WISE} colors for this sample is given in the bottom left panel of Figure\,\ref{fig:WISE2}. The 2D KS test in this case returns $D=0.31$ with the corresponding $p= 0.065$, meaning that at the significance level of $\alpha = 0.05$, the {\it WISE} colors for both samples of DDRGs and GPS/CSOs considered here, were drawn from the same distribution.

Finally, we inspect the {\it WISE} color distribution of our sample of GPS/CSOs against a larger sample of double-lobed radio sources, compiled by \citet{VanVelzen15} by means of an automated search algorithm applied to the FIRST survey. With the adopted five quality cuts --- optimized with respect to the angular size of the respective systems, their radio fluxes, as well as their radio core prominence --- the sample of \citeauthor{VanVelzen15} consists of $24,973$ sources that could be safely considered as representing FR\,II-type radio galaxies, with radio fluxes extending down to the adopted 12\,mJy limit. We performed the same two quality cuts to remove any uncertain measurements as the \citeauthor{Asmus14} sample. We removed all sources that had a null or $<1$ signal to noise ratios (snr) for bands W1, W2, $\&$ W3 as well as all sources with $c\_flags$ not equal to 0. These cuts left us with a sample of $9,740$ sources from the original $24,973$ sources in the \citeauthor{VanVelzen15} sample. The distribution of the {\it WISE} colors for this sample is given in the bottom right panel of Figure\,\ref{fig:WISE2}. The 2D KS for this and our sample returns a value of $D=0.37$ with the corresponding $p=0.0028$, signifying that the two data sets were not drawn from the same distribution. Indeed, it is apparent that one of the main differences here is the lack of GPS/CSOs within the region of the {\it WISE} color diagram occupied by galaxies forming stars at very high rates (starbursts, luminous- and ultraluminous-infrared galaxies), where, in contrast, many double-lobed FIRST radio sources can be found. This could signify that, either the youngest radio sources avoid hosts with very vigorous starformation, \emph{or} that we are dealing with an observational bias that does not allow for differentiation of compact jets and lobes (and therefore for the ``GPS/CSO'' classification) in the presence of a strong radio emission from the central starburst region.

The results from all {\it WISE} color 2D KS tests are summarized in Table\,\ref{tab:KSTest}.

\begin{figure*}[t!]
\centering
\includegraphics[width=0.7\textwidth]{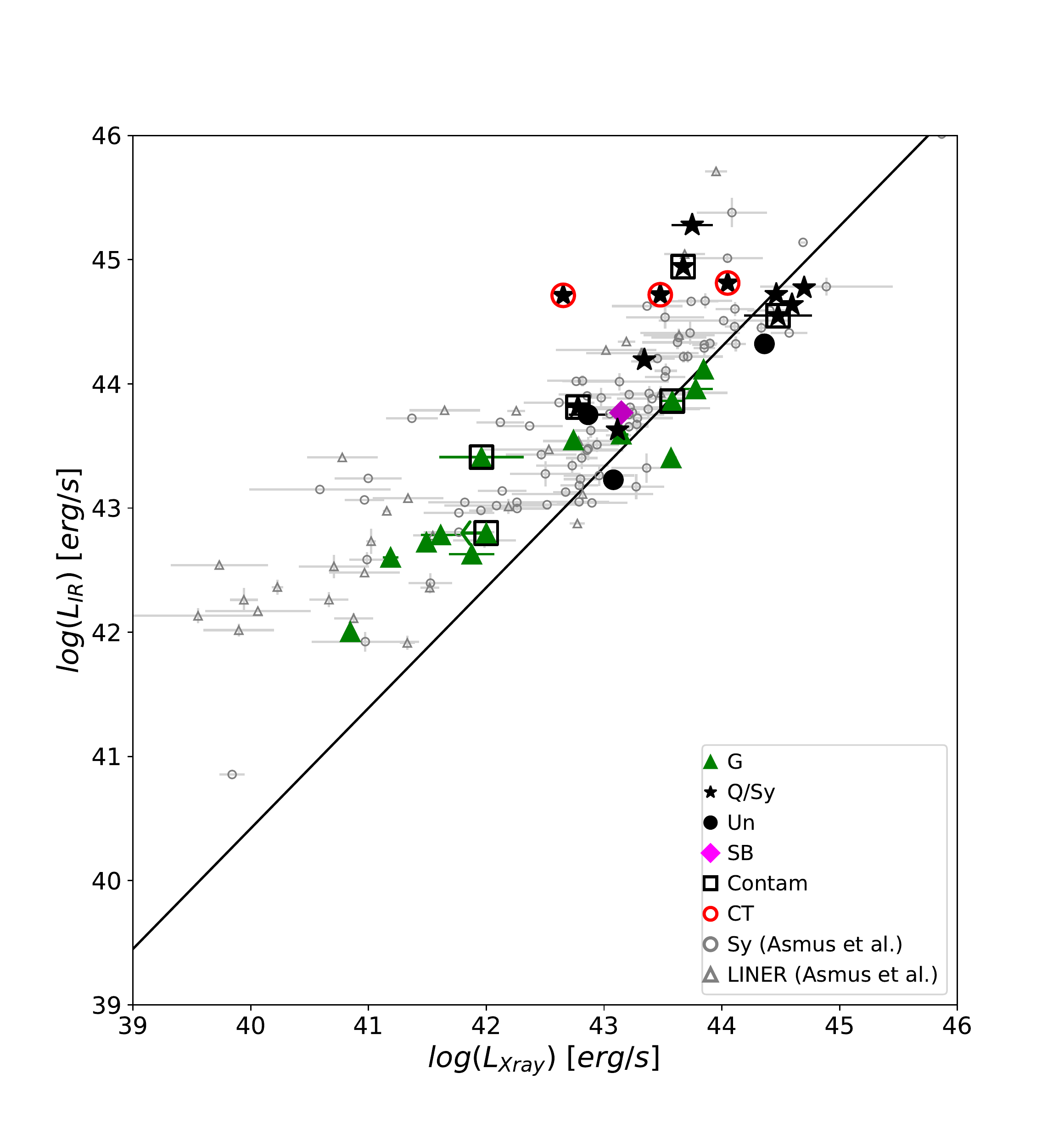}
\includegraphics[width=0.4\textwidth]{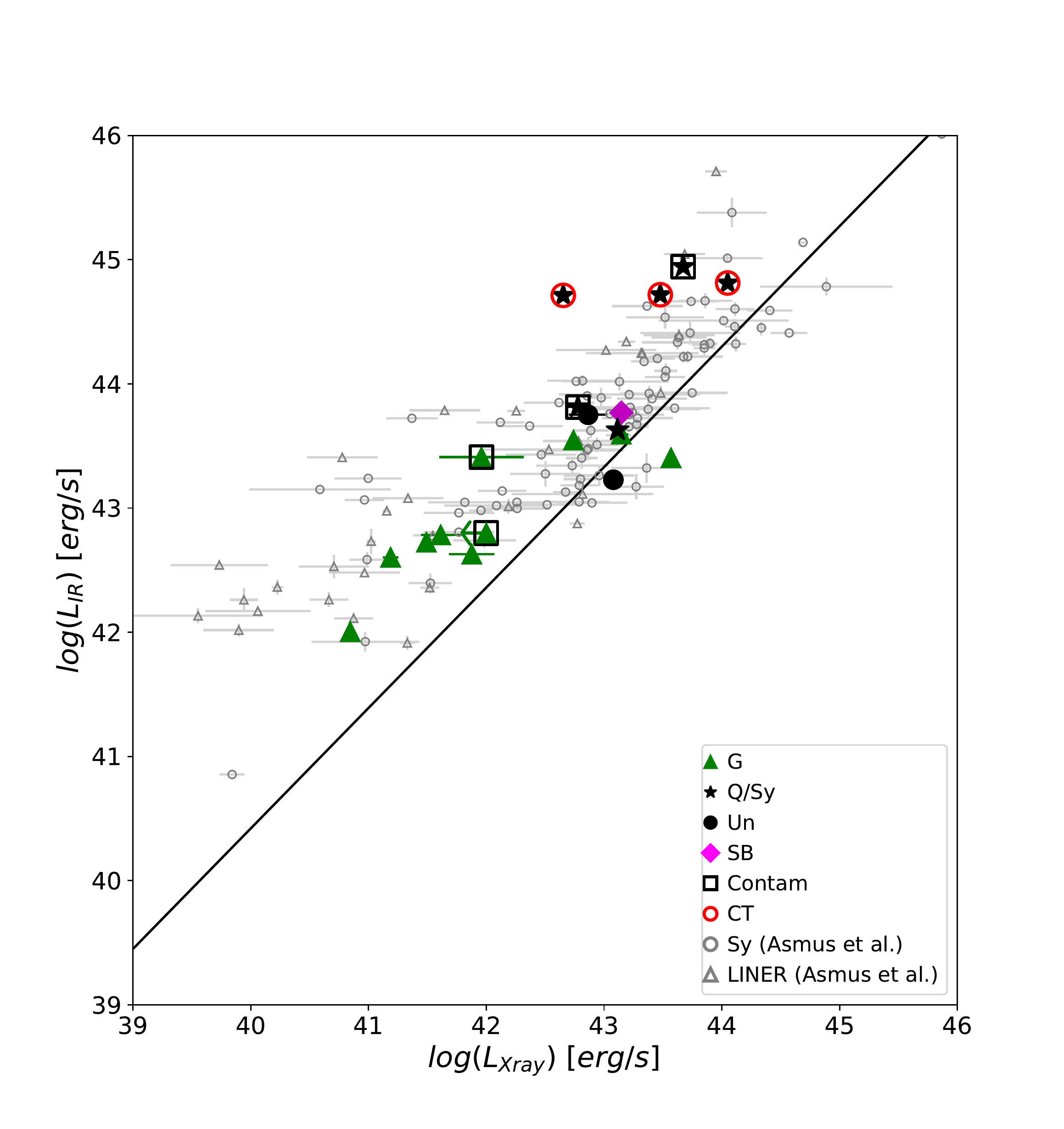}
\includegraphics[width=0.4\textwidth]{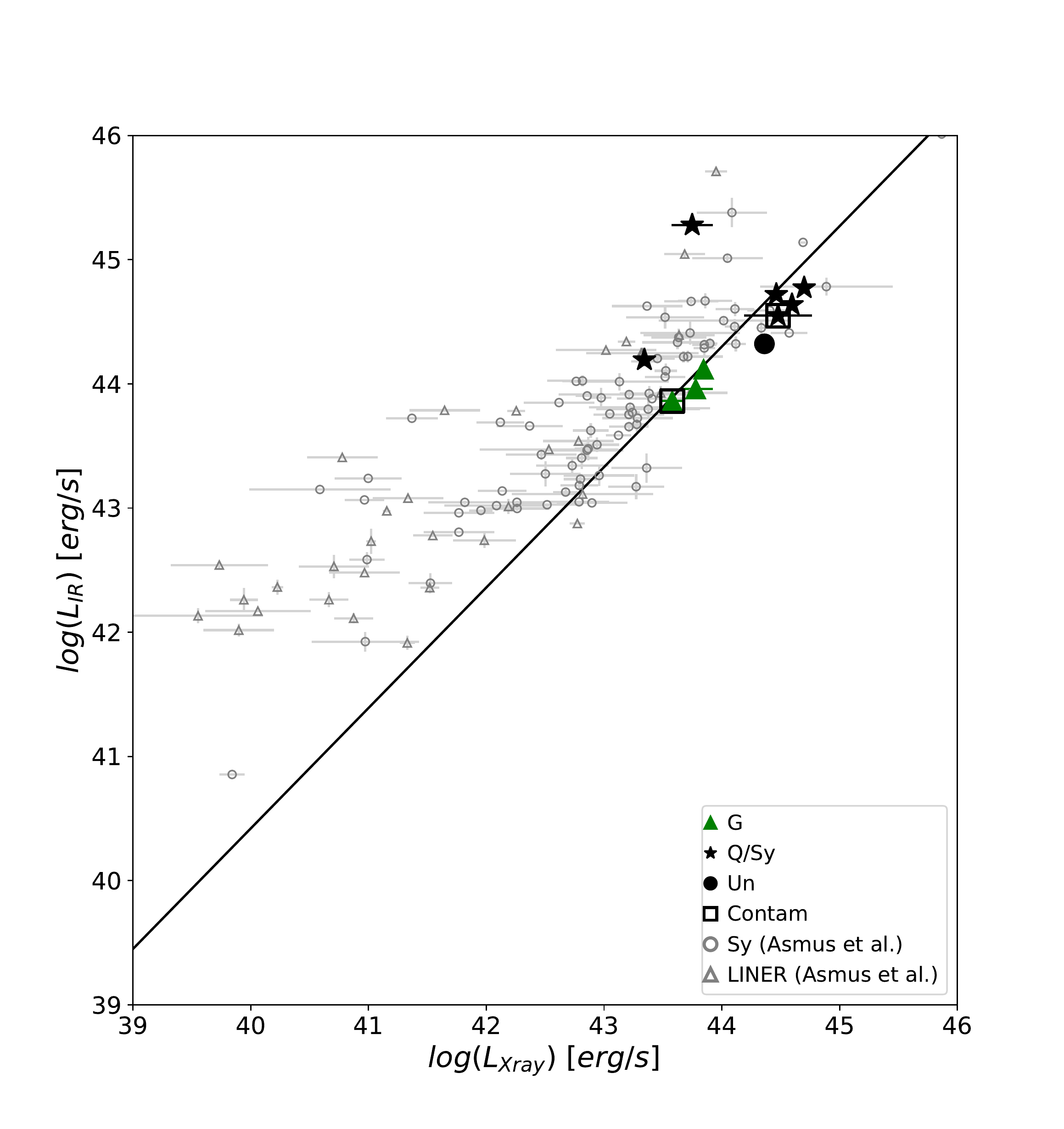}
\caption{Relation of the {\it WISE} 12\,$\mu$m and the intrinsic (absorption-corrected) 2--10\,keV luminosities for our compact radio galaxies denoted by large symbols (symbols and colors are the same as in Figure\,\ref{fig:WISE}). Source $0116+319$ has an upper limit X-ray luminosity and is marked with a left pointing arrow, visible on the lower left end of the plot. The solid line shows the correlation established for nearby AGN by \citet{Asmus15} based on the high-angular resolution MIR (ground-based) observations; small grey symbols denote the AGN from the \citet{Asmus14} sample with {\it IRAS} 12\,$\mu$m luminosities. In the upper panel we show the entire GPS/CSOs sample, in the lower-left panel only the young radio galaxies with $z<0.4$, and in the lower-right panel only those with $z>0.4$.}
\label{fig:MIRX}
\end{figure*}

\begin{figure*}[t!]
\centering
\includegraphics[width=\columnwidth]{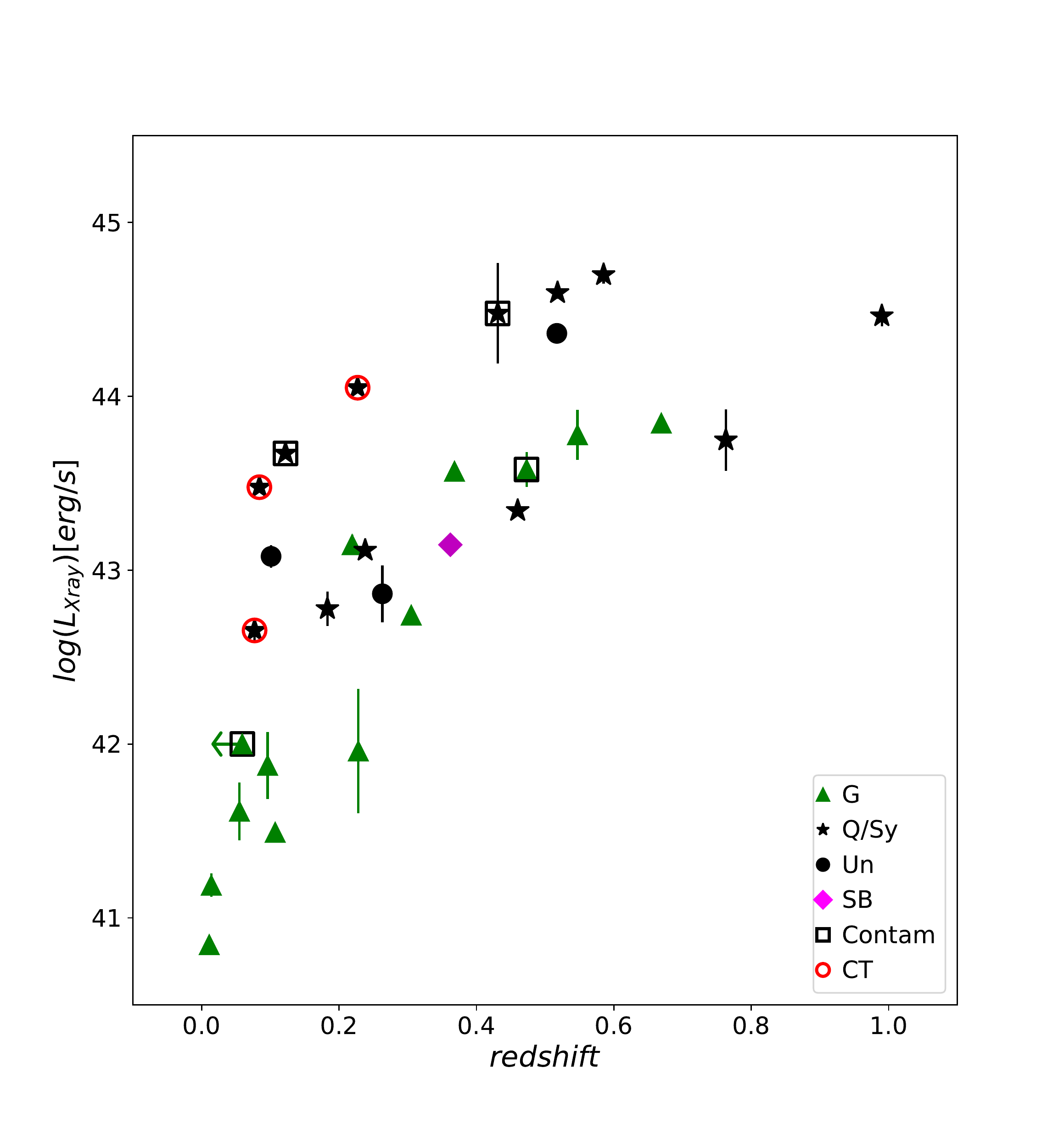}
\includegraphics[width=\columnwidth]{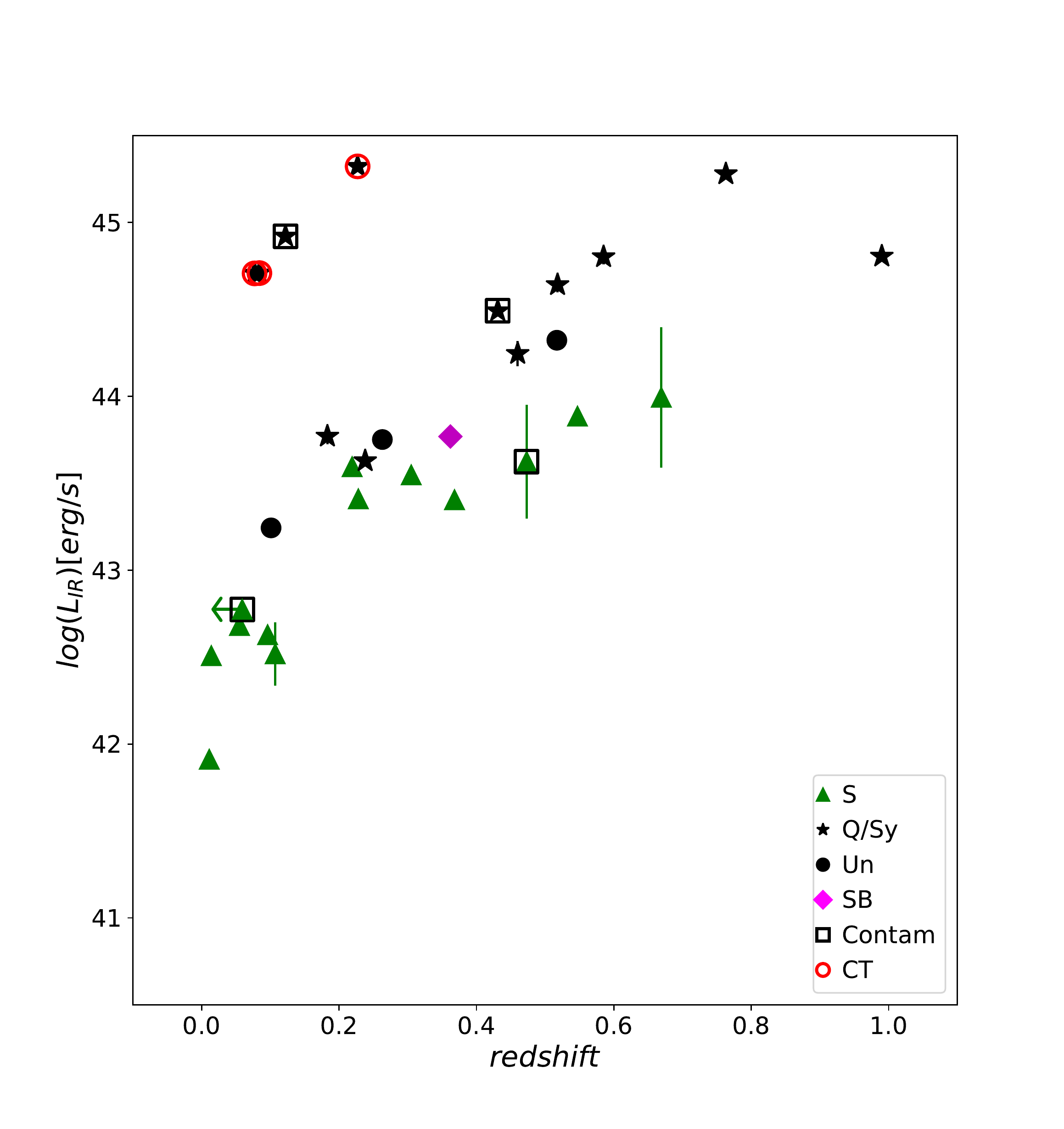}
\caption{The redshift distribution of the intrinsic (absorption-corrected) 2--10\,keV luminosities and the {\it WISE} 12\,$\mu$m luminosities of our sample of GPS/CSOs (left and right panels, respectively). Symbols and colors are the same as in Figure\,\ref{fig:WISE}.}
\label{fig:redshift}
\end{figure*}

\begin{deluxetable}{llc}[t!]
\tabletypesize{\normalsize}
\tablecaption{{\it WISE} 2D KS Test Results}
\tablewidth{0.45\textwidth}
\tablehead{ \colhead{Sample Set}  & \colhead{$p$-value} & \colhead{$D$ value}}
\startdata
 Asmus et al. & $0.0795$ & $0.282$\\
Koziel \& Stasinska  & $0.95 \times 10^{-7}$ & $0.60$ \\
Kuzmicz et al.  & $0.065$ & $0.31$ \\
 Van Velzen et al.   & $0.0028$ & $0.37$
\enddata
\tablenotetext{}{}
\label{tab:KSTest}
\end{deluxetable}

\subsection{MIR/X-ray Correlation}
\label{sec:MIR-X}

A similarity in the MIR colors between our GPS/CSOs and the general population of bright local AGN, prompted us to investigate a correlation between MIR and X-ray luminosities in young radio galaxies. The relationship between MIR and X-ray luminosities in various samples of AGN and in different photon energy ranges have been studied extensively in recent years. Different infrared instruments have been utilized in such analysis, including the low-resolution ISO \citep{Lutz04,Ramos07}, {\it Spitzer} \citep{Hardcastle09,Sazonov12}, AKARI \citep{Matsuta12,Ichikawa12}, and {\it WISE} \citep{Mateos15,Stern15} instruments. Studies with high-resolution ground-based telescopes turned out particularly conclusive in this context, as they disclosed a tight correlation between the \emph{nuclear} MIR luminosities and the 2--10\,keV luminosities of the targeted AGN, which has important implications for understanding the structure of circumnuclear dusty tori in active galaxies in general \citep[see, e.g.,][]{Horst08,Levenson09,Gandhi09,Asmus11,Asmus15}.

In the upper panel of Figure\,\ref{fig:MIRX} we plot the {\it IRAS} 12\,$\mu$m and the intrinsic (absorption-corrected) 2--10\,keV luminosities for the entire sample of our compact radio galaxies (see Table\,\ref{table:1}; large symbols in the figure), along with the comparison AGN sample by \citet{Asmus14} introduced in the previous section, for which the 12\,$\mu$m luminosities were obtained similarly with the {\it IRAS} satellite (small grey symbols). In the cases of our sources we utilize the {\it WISE} 12\,$\mu$m measurements; note the $\sim 6.5''$ {\it WISE} resolution at 12\,$\mu$m compared to the {\it IRAS} angular resolution of $\sim 30''$ at 12\,$\mu$m. The differences in angular resolution of the instruments can result in the {\it WISE} measurements viewing a smaller region of the targeted source. Still, even with such, we are not able to extract solely the innermost nuclear regions of the galaxies, and so all of the sources included in the figure are expected to have a significant contribution from the ISM component in their MIR luminosities, this is in addition to the circumnuclear dust emission (if present). In this context, the solid line in Figure \ref{fig:MIRX} denotes the best-fit linear relation between the logarithms of the luminosities, obtained by \citet{Asmus15} when using the \emph{high-resolution} MIR observations with ground-based telescopes, for their \citep{Asmus14} Seyfert+LINER sample,
\begin{equation}
\log\!\left(\frac{L_{\rm 12\,\mu m}}{10^{43}\,{\rm erg\,s^{-1}}}\right) = 0.33 + 0.97 \, \log\!\left(\frac{L_{\rm 2-10\,keV}}{10^{43}\,{\rm erg\,s^{-1}}}\right) \, .
\end{equation}
As evident from the figure, the AGN from the comparison sample are located mostly above the best-fit correlation line, as expected keeping in mind the relatively poor angular resolution of {\it IRAS}, and the fact that the best-fit correlation was established based on sub-arcsec MIR photometry, and so without any significant ISM emission component. Note that the relative contribution of the ISM to the total MIR flux detected with {\it IRAS} increases for low-luminosity AGN (LINER), as expected.

The distribution of young radio galaxies in the diagram, at the first glance, resembles the distribution of the comparison local AGN sample. The emerging approximate linear scaling between the MIR and X-ray luminosities, along the correlation established by \citet{Asmus15}, could be therefore considered as an indication for the nuclear (disk corona) origin of the observed X-ray fluxes also in the GPS/CSOs sample. Indeed, at least in the case of the confirmed Compton-thick objects, this is the most plausible explanation \citep{Guainazzi04,Siemiginowska16,Sobolewska19a,Sobolewska19b}. On the other hand, in the framework of the model presented in \citet{Stawarz08} and \citet{Ostorero10}, where the X-ray radiative output of GPS/CSOs is dominated by the non-thermal emission of the compact radio lobes, one could expect some correlation between the MIR and X-ray luminosities, because in this model X-ray photons are generated predominantly by the inverse-Comptonization of the circumnuclear dust emission by relativistic electrons within compact lobes; we analyze this possibility in more detail in Appendix\,\ref{appendix} below.

On the other hand, an approximately linear luminosity-luminosity scaling is expected in any flux limited sample of cosmologically distant sources, even in the case of no intrinsic correlations between radiative outputs in various bands \citep[e.g., see the discussion in][]{Singal19}. Neither our GPS/CSOs sample nor the AGN sample by \citet{Asmus14} are, strictly speaking, flux-limited (or complete, in that matter). However, the luminosity-redshift distributions of young radio galaxies analyzed here, presented in Figure\,\ref{fig:redshift}, indicate that we \emph{are} seriously affected by flux limits in both MIR and X-ray bands, and as a result the observed luminosity-luminosity correlation may be simply due to the fact that higher-$z$ sources have to be more luminous in either band, in order to be included in the sample. 

In the lower panels of Figure\,\ref{fig:MIRX}, we plot again the low-resolution {\it WISE} 12\,$\mu$m and the intrinsic (absorption-corrected) 2--10\,keV luminosities for our GPS/CSOs, but divided into the redshift bins $z<0.4$ and $z>0.4$ (left and right panels, respectively). As shown, the overall scalings are still preserved, although there is a significant difference in the luminosity ratio between the two considered redshift bins. We perform the 2D KS test for the $L_{\rm 2-10\,keV} - L_{\rm 12\,\mu m}$ distributions of our GPS/CSOs and the comparison AGN from \citet{Asmus14}, and obtain $D=0.26$ with $p=0.1$ for the entire sample of young radio galaxies, $D=0.26$ with $p=0.25$ when only young radio galaxies at redshifts $z<0.4$ are considered, and $D=0.7$ with $p=1.8 \times 10^{-4}$ for the GPS/CSOs with $z>0.4$ (see Table \ref{tab:LumTest}). These values imply that our null hypothesis --- stating that, in the ``low-resolution MIR luminosity vs. absorption-corrected X-ray luminosity'' representation, the GPS/CSOs and the comparison AGN samples are drawn from the same distribution --- can be rejected only for high-redshift sources.

\begin{deluxetable}{llc}[t!]
\tabletypesize{\normalsize}
\tablecaption{IR--Xray 2D KS Test Results}
\tablewidth{0.45\textwidth}
\tablehead{ \colhead{Sample Set}  & \colhead{$p$-value} & \colhead{$D$ value}}
\startdata
Full Sample & $0.10$ & $0.26$\\
$z<0.4$ & $0.25$ & $0.26$ \\
$z>0.4$ & $1.8 \times 10^{-4}$ & $0.7$
\enddata
\tablenotetext{}{}
\label{tab:LumTest}
\end{deluxetable}

Those high-redshift targets, on the other hand, appear (visually) to match very well the best-fit correlation line of \citet{Asmus15} established based on sub-arcsec MIR photometry of local AGN. This should not be surprising, however, noting that at high MIR luminosities characterizing our $z>0.4$ sources, $L_{\rm 12\,\mu m} \geq 3 \times 10^{43}$\,erg\,s$^{-1}$, any contribution from the ISM component with a moderate star formation rate (up to $\sim 10 M_{\odot}$\,yr$^{-1}$), is expected to be rather minor, and so the fluxes extracted even with low-resolution instruments such as {\it IRAS} or {\it WISE}, should be dominated by the radiative output of circumnuclear dust.

On the related note, we comment that in the case of $z<0.4$ young radio galaxies, the only ones with $L_{\rm 12\,\mu m}$ exceeding $10^{44}$\,erg\,s$^{-1}$, are the three Compton-thick objects 1404+286, 1511+0518 and 2021+614, and the peculiar 1345+125, which can also be classified as Ultraluminous Infrared Galaxies \citep[see][]{Rodriguez09}. Those four sources appear in fact over-luminous in MIR for their given intrinsic (absorption-corrected) X-ray luminosities when compared to the general local AGN population, as well as other local GPS/CSOs (see in this context the right panel in Figure\,\ref{fig:redshift}).

\subsection{Gamma-ray Emitters 1718--649 \& 1146+596}
\label{sec:gamma}

1718--649 and 1146+596 are the two lowest-luminosity sources in our sample, which are rather unique objects for several reasons. 1718--649 is particularly young and compact \citep[LS\,$\simeq 2$\,pc, $\tau \simeq 100$\,yr;][]{Tingay97,Tingay03,Giroletti09}, located at the distance of 60\,Mpc only, and most importantly, it is the first bona-fide CSO detected in the high-energy $\gamma$-ray range with {\it Fermi} Large Area Telescope \citep[LAT; see the analysis and the discussion in][]{Migliori16}. 1146+596 is characterized by a much larger LS\,$\simeq 933$\,pc \citep{Taylor98,Perlman01}\footnote{We note that 1146+596 is characterized by multiple radio outbursts, and the most compact radio component in the source, for which the kinematic age has been estimated as $\sim 60$\,yr, has a linear size of 3.5\,pc \citep{Principe20}.}, but located at a comparable distance of 47\,Mpc, and recently associated with the $\gamma$-ray source 4FGL\,J1149.0+5924 in the {\it Fermi} LAT 8-Year Point Source Catalog \citep[4FGL;][]{4FGL}. The question follows if the observed $\gamma$-ray emission in both sources is related to the jet activity and compact radio lobes in particular \citep{Stawarz08,Kino09}, or is instead due to the star formation activity within the hosts, analogous to the LAT-detected, star-forming and starburst galaxies \citep{Ackermann12,Hayashida13,Rojas16}. 

In order to address this question, we have collected all archival infrared data for both sources, including near-infrared 2MASS and MIR {\it Spitzer} spectra, as well as mm fluxes from WMAP, and estimated the integrated $8-1000$\,$\mu$m luminosities, and next the star-formation rates assuming the \citet{Kennicutt98} scaling relation 
\begin{equation}
\frac{{\rm SFR}}{M_{\odot} \, {\rm yr^{-1}}} =   1.7 \, \epsilon \times \frac{L_{8-1000\,\mu {\rm m}}}{ 10^{10} \, L_{\odot}} ,
\end{equation}
with the initial mass function factor $\epsilon = 0.79$ \citep{Ackermann12}. The resulting values, given in Table\,\ref{tab:SFR}, are consistent with the SFR estimates by \citet{Willett10} for 1718--649, based on either Ne lines or PAH features in the {\it Spitzer} high-resolution spectra (see the Table). In the case of 1146+596 we achieve a lower SFR estimate based on the integrated luminosity, which may be due in part to poor data quality (i.e. lack of data points over the infrared range) as well as due to the fact that some MIR emission can come from the dust heated by old stellar populations and not indeed by star formation itself. As both sources are the weakest MIR emitters in the sample, the SFR at the corresponding level of $\sim 1\,M_{\odot}$\,yr$^{-1}$ should be considered as the limiting value for young radio galaxies analyzed here. Assuming further the best-fit correlation derived by \citet{Ackermann12} for the LAT-detected, star-forming and starburst galaxies,
\begin{equation}
\log \left(\frac{L_{0.1-100\,{\rm GeV}}^{\rm [ISM]}}{10^{39} \,{\rm erg\,s^{-1}}} \right) = 1.17 \, \log \left(\frac{L_{8-1000\,\mu {\rm m}}}{10^{10} L_{\odot}}\right) + 0.28 ,
\end{equation}
we next derived the ISM-related $0.1-100$\,GeV luminosities for the two targets, which are both at the level of $\sim 10^{38-39}$\,erg\,s$^{-1}$, as listed in Table\,\ref{tab:SFR}.

\begin{deluxetable*}{ccccccc}[t!]
\tabletypesize{\normalsize}
\tablecaption{SFRs, IR luminosities, and $\gamma$-ray luminosties of 1718--649 \& 1146+596}
\tablewidth{0pt}
\tablehead{
\colhead{Name} & \colhead{$L_{8-1000\mu m}$} & \colhead{SFR$_{\rm IR}$} & \colhead{SFR$_{\rm Ne}$} & \colhead{SFR$_{\rm PAH}$} &  \colhead{$L_{0.1-100\,{\rm GeV}}^{\rm [ISM]}$} & \colhead{$L_{0.1-100\,{\rm GeV}}$} \\
\colhead{~~} &  \colhead{erg\,s$^{-1}$} & \colhead{$M_{\odot}$\,yr$^{-1}$} & \colhead{$M_{\odot}$\,yr$^{-1}$}&  \colhead{$M_{\odot}$\,yr$^{-1}$} &\colhead{erg\,s$^{-1}$} & \colhead{erg\,s$^{-1}$} \\
\colhead{(1)} & \colhead{(2)} & \colhead{(3)} & \colhead{(4)} & \colhead{(5)} &  \colhead{(6)} & \colhead{(7)} 
}
\startdata
1146+596  & $5.5 \times 10^{42}$ & $\sim 0.13$ & $0.5 \pm 0.1 $ & $\sim 0.3$ & $\sim 2 \times 10^{38}$ & $\sim 0.6 \times 10^{42}$\\
1718--649  &  $3.7 \times 10^{43}$ & $\sim0.89$ & $1.8 \pm 0.1$ & $\sim 0.8$ &  $\sim 2 \times 10^{39}$ & $\sim 2 \times 10^{42}$
\enddata
\tablenotetext{}{
{\bf col(1)} --- Name of the source; 
{\bf col(2)} --- Integrated infrared luminosity derived from archival $8-1000$\,$\mu$m data; 
{\bf col(3)} --- Star formation rate for the given integrated infrared luminosity in col(2);
{\bf col(4)} --- Star formation rate calculated from Ne emission lines in \citet{Willett10}; 
{\bf col(5)} --- Star formation rate calculated from Polycyclic aromatic hydrocarbon (PAH) emission lines in \citet{Willett10}; 
{\bf col(6)} --- $0.1-100$\,GeV luminosity estimated emission based on the integrated infrared flux in col(2); 
{\bf col(7)} --- The observed {\it Fermi}-LAT $0.1-100$\,GeV luminosity. 
}
\label{tab:SFR}
\end{deluxetable*}

Meanwhile, the integrated $0.1-100$\,GeV photon flux of the high-energy $\gamma$-ray counterpart of 1718--649 provided by \citet{Migliori16}, $(11.5\pm 0.3)\times 10^{-9}$\,ph\,cm$^{-2}$\,s$^{-1}$, along with the best-fit photon index $\Gamma_{\gamma} = 2.9\pm0.3$, yields the luminosity $L_{0.1-100\,{\rm GeV}} \sim 2 \times 10^{42}$\,erg\,s$^{-1}$, which is three orders of magnitude larger than the one expected for the ISM emission as estimated above. Similarly, for 1146+596, based on the integrated $1-100$\,GeV photon flux of the high-energy $\gamma$-ray counterpart provided in 4FGL, $(2.17\pm 0.35)\times 10^{-10}$\,ph\,cm$^{-2}$\,s$^{-1}$, along with the best-fit photon index $\Gamma_{\gamma} = 2.06\pm0.11$, we derive the luminosity $L_{0.1-100\,{\rm GeV}} \sim 6  \times 10^{41}$\,erg\,s$^{-1}$, which is again orders of magnitude larger than the  expected emission for the ISM component alone. 

Hence, we conclude that the $\gamma$-ray emission detected with {\it Fermi}-LAT from 1718--649 and 1146+596, is undoubtedly related to the jet activity in the sources, and not the star formation activity. This strengthens the expectation that compact lobes/jets in young radio galaxies are in general $\gamma$-ray emitters. Moreover, we note that the two CSOs detected by {\it Fermi} LAT so far, are in fact the \emph{nearest} objects in our sample. This suggests that with a longer accumulation of the LAT data, the most promising candidates for the $\gamma$-ray detection should be 0116+319 and 0402+379 (luminosity distances $\sim 250$\,Mpc; see Table\,\ref{table:1}), and --- even more excitingly --- the two Compton-thick CSOs 1404+286 and 1511+0518 (luminosity distances $\sim 350$\,Mpc).

The existence of an intrinsic link between the X-ray and $\gamma$-ray luminosities among CSO/GPS galaxies, is still an open question, having only two such sources detected in high-energy gamma-rays. However, theoretical modeling suggest there may be a linear correlation between the X-ray and $\gamma$-ray bands related to the radiative output of the jets in young radio sources, for further discussion see \citet{Migliori14}. Additional $\gamma$-ray detections of young radio galaxies could shed light on this issue and allow more conclusive statements on this relationship.

\section{Conclusions} \label{sec:conclusions}

In this paper we have discussed the MIR properties of the most compact (GPS/CSO-type) radio galaxies, based predominantly on the lower-resolution MIR data provided by {\it WISE} and {\it IRAS} satellites (augmented in a few cases by the {\it Spitzer} observations). We have restricted our analysis to only those objects which have been observed in X-rays with either XMM-{\it Newton} or {\it Chandra}, and this resulted in the sample of 29 objects.

The inspection of optical images for the targets, implies that the analyzed GPS/CSOs are hosted predominantly by red/yellow galaxies with an elliptical shape, with several cases containing distorted morphology or signs of galaxy interaction. 

In our sample we found a variety of {\it WISE} colors, suggesting that the MIR continua of the studied sources are contributed not exclusively by the circumnuclear dust, but also by the ISM of host galaxies, and in a few cases even by the non-thermal emission of compact jets and lobes. Such a diversity resembles a general population of bright local AGN \citep{Asmus14}. In particular, we found that young radio galaxies with the MIR emission dominated by the ISM component, populate almost uniformly the entire region occupied by galaxies (from elliptical to starbursts) with a wide range of a pronounced star formation activity (rates $\geq 0.5\,M_{\odot}$\,yr$^{-1}$). This constitutes a significant difference with the population of the evolved ordinary FR\,II radio galaxies \citep{Koziel11}, which are clustered within the area of the MIR color diagram occupied by elliptical galaxies with very little star formation activity. This unexpected result --- keeping in mind the general expectation that luminous CSOs are young progenitor of FR\,II radio galaxies --- could signal a negative feedback at work, meaning that that expanding jets and lobes, during their evolution from a compact CSO phase up to the evolved classical double phase, suppress the star formation in galactic hosts. Yet we also found that the distribution of the MIR colors in our GPS/CSOs sample is statistically indistinguishable from the distribution of MIR colors of large-scale radio galaxies but with recurrent jet activity \citep[DDRGs;][]{Kuzmicz17}. We quantified our statement by performing two-dimensional Kolmogorov-Smirnov tests.

All these findings seems therefore to imply that (1) triggering radio jets in AGN does not differentiate between elliptical hosts with substantially different fractions of young stars, and (2) it is the jet duty cycle --- and not the jet launching itself --- which is related to the star formation rate within the host, in that radio galaxies hosted by galaxies with more pronounced population of young stars are either typically short lived (i.e., not surviving long enough to form an extended FR\,II structure), or characterized by a highly modulated/recurrent jet activity. We note in this context that, in the accompanying paper \citet{Wojtowicz19}, we derived the bolometric luminosities of the accretion disks and the black hole masses for about half of the GPS/CSOs sample analyzed here, and demonstrated that the corresponding accretion rates are in all the cases high, namely between $1\%$ and $20\%$ in Eddington units.

The caveat here is the {\it WISE} color distribution of our sample of GPS/CSOs against a larger sample of double-lobed radio sources from \citet{VanVelzen15}, suggest at the same time that the youngest radio sources avoid hosts with very vigorous starformation. However, a more plausible explanations is that there exists an observational bias which does not allow for differentiation of compact jets and lobes (and therefore for the ``GPS/CSO'' classification) in the presence of a strong radio emission from the central starburst region.

The distribution of the sub-sample of our sources with $z<0.4$ on the low-resolution MIR vs. absorption-corrected X-ray luminosity plane is consistent with the distribution of a sample of local AGN on the same plane \citep{Asmus14}. High-resolution (sub-arcsec) MIR photometry of the studied sources, along with deep X-ray exposures, are required in order to make any conclusive statements on the origin of the emerging luminosity-luminosity correlation.

Interestingly, all the confirmed Compton-thick objects in the sample --- 1404+286, 1511+0518 and 2021+614 --- are characterized by the {\it WISE} colors consistent with quasar identification, and, in addition, appear over-luminous in MIR for the given level of their intrinsic X-ray emission when compared to the general local AGN population as well as other local GPS/CSOs. Hence, {\it WISE} color diagnostics and photometry could, in general, be considered as a very useful tool when selecting candidates for Compton-thick AGN among compact radio galaxies. Note in this context that, compact radio sources are particularly abundant in flux-limited radio samples, constituting $\sim 10\%$ of the bright radio-source population \citep{Odea98,Sadler16}. 

Finally, we discussed in more detail the particular case of the $\gamma$-ray--detected 1718--649 and 1146+596. By means of a comparison with the sample of star-forming and starburst galaxies detected with {\it Fermi}-LAT, we have argued that the $\gamma$-ray flux observed from these, is undoubtedly related to the jet activity in the sources, and not due to the star formation activity. This strengthens the expectation that compact lobes/jets in young radio galaxies are in general $\gamma$-ray emitters, albeit relatively weak so that the {\it Fermi}-LAT detections are currently limited to the two the nearest objects in the sample. 

\acknowledgments

This work was supported by the Fulbright Program and in collaboration with the Astronomical Observatory of the Jagiellonian University. EK, \L S, AW, and VM were supported by Polish NSC grant 2016/22/E/ST9/00061. L.O. acknowledges partial support from the INFNGrant InDark and the grant of the Italian Ministry of Education, University and Research (MIUR) (L.232/2016)“ECCELLENZA1822 D206 - Dipartimento di Eccellenza2018-2022 Fisica” awarded to the Dept. of Physics of the University of Torino. M.S. and A.S. were supported by NASA contract NAS8-03060 (Chandra X-ray Center). Work at the Naval Research Laboratory is supported by the Chief of Naval Research.

The authors thank D. Asmus, P. Gandhi, and A. Ku{\'z}micz for the discussion and comments.

This publication makes use of data products from the Wide-field Infrared Survey Explorer, which is a joint project of the University of California, Los Angeles, and the Jet Propulsion Laboratory/California Institute of Technology, funded by the National Aeronautics and Space Administration.

This research has made use of the NASA/ IPAC Infrared Science Archive, which is operated by the Jet Propulsion Laboratory, California Institute of Technology, under contract with the National Aeronautics and Space Administration.

Funding for the Sloan Digital Sky Survey IV has been provided by the Alfred P. Sloan Foundation, the U.S. Department of Energy Office of Science, and the Participating Institutions. SDSS acknowledges support and resources from the Center for High-Performance Computing at the University of Utah. The SDSS web site is \url{www.sdss.org} .

The Pan-STARRS1 Surveys (PS1) and the PS1 public science archive have been made possible through contributions by the Institute for Astronomy, the University of Hawaii, the Pan-STARRS Project Office, the Max-Planck Society and its participating institutes, the Max Planck Institute for Astronomy, Heidelberg and the Max Planck Institute for Extraterrestrial Physics, Garching, The Johns Hopkins University, Durham University, the University of Edinburgh, the Queen's University Belfast, the Harvard-Smithsonian Center for Astrophysics, the Las Cumbres Observatory Global Telescope Network Incorporated, the National Central University of Taiwan, the Space Telescope Science Institute, the National Aeronautics and Space Administration under Grant No. NNX08AR22G issued through the Planetary Science Division of the NASA Science Mission Directorate, the National Science Foundation Grant No. AST-1238877, the University of Maryland, Eotvos Lorand University (ELTE), the Los Alamos National Laboratory, and the Gordon and Betty Moore Foundation.

\appendix

\section{Non-thermal X-ray Emission of Young Radio Galaxies}
\label{appendix}

Below we provide a very simplified approximation for the expected X-ray luminosity of compact radio lobes in compact radio galaxies, following \citet{Stawarz08}. In this model, the monochromatic lobes' luminosity is
\begin{equation}
\varepsilon L_{\varepsilon} \simeq \frac{2}{3} c \sigma_{\rm T} V U_{\rm MIR} \, \left[\gamma^3 N_{\rm e}\!(\gamma)\right]_{\gamma=\sqrt{\varepsilon/\varepsilon_0}} \, ,
\end{equation}
where $V$ is the volume of the lobes, $U_{\rm MIR}$ is the energy density of the circumnuclear dust emission at the position of the lobes, $N_{\rm e}\!(\gamma)$ is the energy spectrum of ultrarelativistic electrons with Lorentz factors $\gamma$, and $\varepsilon_0$ is the energy of the Compton-upscattered seed photon. By following the evolution of the electrons subjected to radiative and adiabatic energy losses within the expanding lobes, \citet[section 3.1. therein]{Stawarz08} showed that in the case of a single power-law injection rate $Q_{\rm e}(\gamma) = k \, \gamma^{-s}$ with $s \geq 2.0$ and the normalization constant $k$, for a given age of a source $\tau$ the electron spectrum can be approximated by $N_{\rm e}\!(\gamma) \simeq \tau \, (k/V) \times f(\gamma)$, where $f(\gamma) = \gamma^{-s}$ for $\gamma < \gamma_{\rm cr}$ and $f(\gamma) =  \gamma_{\rm cr} \gamma^{-s-1}$ otherwise, and $ \gamma_{\rm cr} \propto L_{\rm j}^{-1/2}$ is the cooling break energy depending on the jet's total kinetic luminosity $L_{\rm j}$. We further assume an energy equipartition between ultrarelativistic electrons and the lobes' magnetic field,  meaning $V \, U_{\rm e} \simeq \tau L_{\rm j}$, where $U_{\rm e} \equiv \int \gamma m_{\rm e}c^2 N_{\rm e}\!(\gamma) d\gamma$ is the electron energy density. Note that $\tau = {\rm LS}/v_{\rm h}$, where LS is the linear size of the lobes, taken as the distance between the the jet termination shocks (hereafter ``hotspots'') in a given source, and $v_{\rm h}$ is the expansion velocity of the lobes, measured as the separation velocity between the hotspots.

For the seed photon's energy density at the position of the lobes, we simply take $U_{\rm MIR} \simeq L_{\rm MIR}/4 \pi R^2 c$, where $R \simeq (\sqrt[3]{3}/8) \, {\rm LS}$ is the effective radius of the lobes, approximated as an ellipsoid with a semi-major axis $a={\rm LS}/2$ and semi-minor axis $b=a/4$ \citep[see in this context][]{Kawakatu08,Wojtowicz19}, and $L_{\rm MIR} \sim 3 \times L_{\rm 12\,\mu m}$ is the total MIR luminosity of the hot dusty torus, with a mean photon energy $\varepsilon_0 = hc/\lambda_0$ for $\lambda_0 \sim 12$\,$\mu$m. Given all the above, one may find
\begin{equation}
\frac{L_{\rm 2-10\,keV}}{L_{\rm 12\,\mu m}} \simeq \frac{32 \, \sigma_{\rm T}}{3^{2/3} \pi m_{\rm e}c^2} \, \frac{L_{\rm j}}{v_{\rm h}  {\rm LS}} \, \frac{\int_{2}^{10} \left[\gamma^3 f\!(\gamma)\right]_{\gamma=\sqrt{\frac{x}{x_0}}}  \frac{dx}{x} }{\int_1^{1e5}  \gamma f\!(\gamma) d\gamma} \, ,
\label{eq:scaling}
\end{equation}
where $x\equiv \varepsilon/{\rm keV}$. We next note, for the injection index $s=2.5$ corresponding to the mean radio spectral index of compact radio galaxies \citep{deVries97}, the last term in the above relation containing integrals over $f(\gamma)$, depends only weakly on the jet kinetic power, and within a wide range of $L_{\rm j}$ reads as $\sim 20$. Hence, a very simple approximate scaling emerges
\begin{equation}
\frac{L_{\rm 2-10\,keV}}{L_{\rm 12\,\mu m}} \sim \left(\frac{L_{\rm j}}{10^{45}\,{\rm erg\,s^{-1}}}\right) \, \left(\frac{{\rm LS}}{{\rm pc}}\right)^{-1} \, \left(\frac{v_{\rm h}}{c}\right)^{-1}
 \, .
\end{equation}

One should keep in mind that the scaling relation derived above relies on several crude approximations, in particular the spectral shape of the electron injection function, so it is possible that in several cases a more detailed modelling, taking into account the exact shape of the radio continua of the studied sources \citep[as presented in][]{Ostorero10}, could result in an elevated level of the expected inverse-Compton emission of compact lobes in the X-ray domain. Additionally, the true jet kinetic luminosities may be larger than the minimum values utilized here. Keeping all these caveats in mind, we conclude that a significant contribution of the non-thermal emission of compact lobes to the observed radiative output of compact radio galaxies in the X-ray domain, remains a plausible option, and that the resulting X-ray luminosity should scale with the MIR luminosity of dusty tori, albeit with a wider scatter due to the dependance of the $L_{\rm 2-10\,keV}/L_{\rm 12\,\mu m}$ ratio on the jet kinetic luminosity $L_{\rm j}$, the linear size of radio lobes, and the expansion velocity of compact lobes $v_{\rm h}$.


\begin{thebibliography}{}

\bibitem[Abolfathi et al.(2018)]{Abolfathi18} Abolfathi, B., Aguado, D.~S., Aguilar, G., et al.\ 2018, \apjs, 235, 42 

\bibitem[Ackermann et al.(2012)]{Ackermann12} Ackermann, M., Ajello, M., Allafort, A., et al.\ 2012, \apj, 755, 164 

\bibitem[Allison et al.(2019)]{Allison19} Allison, J.~R., Mahony, E.~K., Moss, V.~A., et al.\ 2019, \mnras, 482, 2934 

\bibitem[Asmus et al.(2011)]{Asmus11} Asmus, D., Gandhi, P., Smette, A., H{\"o}nig, S.~F., \& Duschl, W.~J.\ 2011, \aap, 536, A36 

\bibitem[Asmus et al.(2014)]{Asmus14} Asmus, D., H{\"o}nig, S.~F., Gandhi, P., Smette, A., \& Duschl, W.~J.\ 2014, \mnras, 439, 1648 

\bibitem[Asmus et al.(2015)]{Asmus15} Asmus, D., Gandhi, P., H{\"o}nig, S.~F., Smette, A., \& Duschl, W.~J.\ 2015, \mnras, 454, 766 

\bibitem[Axon, et al.(2000)]{Axon00} Axon D.~J., Capetti A., Fanti R., Morganti R., Robinson A., Spencer R., 2000, AJ, 120, 2284

\bibitem[Beuchert et al.(2018)]{Beuchert18} Beuchert, T., Rodr{\'{\i}}guez-Ardila, A., Moss, V.~A., et al.\ 2018, \aap, 612, L4 

\bibitem[Chambers et al.(2016)]{Chambers16} Chambers, K.~C., Magnier, E.~A., Metcalfe, N., et al.\ 2016, arXiv e-prints, arXiv:1612.05560

\bibitem[Czerny \& You(2016)]{Czerny16} Czerny, B., \& You, B.\ 2016, Astronomische Nachrichten, 337, 73 

\bibitem[de Vries et al.(1997)]{deVries97} de Vries, W.~H., Barthel, P.~D., \& O'Dea, C.~P.\ 1997, \aap, 321, 105 

\bibitem[de Vries, et al.(1998)]{deVries98} de Vries W.~H., et al., 1998, ApJ, 503, 138


\bibitem[Dicken et al.(2012)]{Dicken12} Dicken, D., Tadhunter, C., Axon, D., et al.\ 2012, \apj, 745, 172 

\bibitem[Fanti et al.(2000)]{Fanti00} Fanti, C., Pozzi, F., Fanti, R., et al.\ 2000, \aap, 358, 499 

\bibitem[Fasano, \& Franceschini(1987)]{Fasano87} Fasano, G., \& Franceschini, A.\ 1987, \mnras, 225, 155

\bibitem[The Fermi-LAT collaboration(2019)]{4FGL} The Fermi-LAT collaboration\ 2019, arXiv e-prints, arXiv:1902.10045

\bibitem[Gandhi et al.(2009)]{Gandhi09} Gandhi, P., Horst, H., Smette, A., et al.\ 2009, \aap, 502, 457 

\bibitem[Giroletti \& Polatidis(2009)]{Giroletti09} Giroletti, M., \& Polatidis, A.\ 2009, Astronomische Nachrichten, 330, 193 

\bibitem[Glowacki et al.(2017)]{Glowacki17} Glowacki, M., Allison, J.~R., Sadler, E.~M., et al.\ 2017, \mnras, 467, 2766 

\bibitem[Guainazzi et al.(2004)]{Guainazzi04} Guainazzi, M., Siemiginowska, A., Rodriguez-Pascual, P., \& Stanghellini, C.\ 2004, \aap, 421, 461 

\bibitem[Guainazzi et al.(2006)]{Guainazzi06} Guainazzi, M., Siemiginowska, A., Stanghellini, C., et al.\ 2006, \aap, 446, 87 

\bibitem[Hardcastle et al.(2009)]{Hardcastle09} Hardcastle, M.~J., Evans, D.~A., \& Croston, J.~H.\ 2009, \mnras, 396, 1929 

\bibitem[Hayashida et al.(2013)]{Hayashida13} Hayashida, M., Stawarz, {\L}., Cheung, C.~C., et al.\ 2013, \apj, 779, 131 

\bibitem[Heckman et al.(1994)]{Heckman94} Heckman, T.~M., O'Dea, C.~P., Baum, S.~A., \& Laurikainen, E.\ 1994, \apj, 428, 65 

\bibitem[Horst et al.(2008)]{Horst08} Horst, H., Gandhi, P., Smette, A., \& Duschl, W.~J.\ 2008, \aap, 479, 389 

\bibitem[Ichikawa et al.(2012)]{Ichikawa12} Ichikawa, K., Ueda, Y., Terashim
a, Y., et al.\ 2012, \apj, 754, 45 

\bibitem[Jauncey, et al.(1986)]{Jauncey86} Jauncey D.~L., White G.~L., Batty M.~J., Preston R.~A., 1986, AJ, 92, 1036

\bibitem[Jia et al.(2013)]{Jia13} Jia, J., Ptak, A., Heckman, T., \& Zakamska, N.~L.\ 2013, \apj, 777, 27 

\bibitem[Kawakatu et al.(2008)]{Kawakatu08} Kawakatu, N., Nagai, H., \& Kino, M.\ 2008, \apj, 687, 141 

\bibitem[Kellermann et al.(2004)]{Kellermann04} Kellermann, K.~I., Lister, M.~L., Homan, D.~C., et al.\ 2004, \apj, 609, 539 

\bibitem[Kennicutt(1998)]{Kennicutt98} Kennicutt, R.~C., Jr.\ 1998, \apj, 498, 541 

\bibitem[Kino et al.(2009)]{Kino09} Kino, M., Ito, H., Kawakatu, N., \& Nagai, H.\ 2009, \mnras, 395, L43 

\bibitem[Kozie{\l}-Wierzbowska, \& Stasi{\'n}ska(2011)]{Koziel11} Kozie{\l}-Wierzbowska, D., \& Stasi{\'n}ska, G.\ 2011, \mnras, 415, 1013

\bibitem[Kunert-Bajraszewska et al.(2014)]{Kunert14} Kunert-Bajraszewska, M., Labiano, A., Siemiginowska, A., \& Guainazzi, M.\ 2014, \mnras, 437, 3063 

\bibitem[Ku{\'z}micz et al.(2017)]{Kuzmicz17} Ku{\'z}micz, A., Jamrozy, M., Kozie{\l}-Wierzbowska, D., et al.\ 2017, \mnras, 471, 3806

\bibitem[Levenson et al.(2009)]{Levenson09} Levenson, N.~A., Radomski, J.~T., Packham, C., et al.\ 2009, \apj, 703, 390 

\bibitem[Lutz et al.(2004)]{Lutz04} Lutz, D., Maiolino, R., Spoon, H.~W.~W., \& Moorwood, A.~F.~M.\ 2004, \aap, 418, 465 

\bibitem[Massaro et al.(2012)]{Massaro12} Massaro, F., D'Abrusco, R., Tosti, G., et al.\ 2012, \apj, 750, 138 

\bibitem[Mateos et al.(2012)]{Mateos12} Mateos, S., Alonso-Herrero, A., Carrera, F.~J., et al.\ 2012, \mnras, 426, 3271 

\bibitem[Mateos et al.(2015)]{Mateos15} Mateos, S., Carrera, F.~J., Alonso-Herrero, A., et al.\ 2015, \mnras, 449, 1422 

\bibitem[Matsuta et al.(2012)]{Matsuta12} Matsuta, K., Gandhi, P., Dotani, T., et al.\ 2012, \apj, 753, 104 

\bibitem[Migliori et al.(2014)]{Migliori14} Migliori, G., Siemiginowska, A., Kelly, B.~C., et al.\ 2014, \apj, 780, 165 

\bibitem[Migliori et al.(2016)]{Migliori16} Migliori, G., Siemiginowska, A., Sobolewska, M., et al.\ 2016, \apjl, 821, L31 

\bibitem[O'Dea(1998)]{Odea98} O'Dea, C.~P.\ 1998, \pasp, 110, 493 

\bibitem[O'Dea(2016)]{Odea16} O'Dea, C.~P.\ 2016, Astronomische Nachrichten, 337, 141 

\bibitem[Ostorero et al.(2010)]{Ostorero10} Ostorero, L., Moderski, R., Stawarz, {\L}., et al.\ 2010, \apj, 715, 1071 

\bibitem[Ostorero et al.(2017)]{Ostorero17} Ostorero, L., Morganti, R., Diaferio, A., et al.\ 2017, \apj, 849, 34 

\bibitem[Peacock(1983)]{Peacock83} Peacock, J.~A.\ 1983, \mnras, 202, 615

\bibitem[Perlman et al.(2001)]{Perlman01} Perlman, E.~S., Stocke, J.~T., Conway, J., \& Reynolds, C.\ 2001, \aj, 122, 536 

\bibitem[Perucho(2016)]{Perucho16} Perucho, M.\ 2016, Astronomische Nachrichten, 337, 18

\bibitem[Principe et al.(2020)]{Principe20} Principe, G., Migliori, G., Johnson, T.~J., et al.\ 2020, \aap, 635, A185

\bibitem[Ramos Almeida et al.(2007)]{Ramos07} Ramos Almeida, C., P{\'e}rez Garc{\'{\i}}a, A.~M., Acosta-Pulido, J.~A., \& Rodr{\'{\i}}guez Espinosa, J.~M.\ 2007, \aj, 134, 2006 

\bibitem[Rodr{\'{\i}}guez Zaur{\'{\i}}n et al.(2009)]{Rodriguez09} Rodr{\'{\i}}guez Zaur{\'{\i}}n, J., Tadhunter, C.~N., \& Gonz{\'a}lez Delgado, R.~M.\ 2009, \mnras, 400, 1139 

\bibitem[Rojas-Bravo \& Araya(2016)]{Rojas16} Rojas-Bravo, C., \& Araya, M.\ 2016, \mnras, 463, 1068 

\bibitem[Romani et al.(2014)]{Romani14} Romani, R.~W., Forman, W.~R., Jones, C., et al.\ 2014, \apj, 780, 149 

\bibitem[Sadler(2016)]{Sadler16} Sadler, E.~M.\ 2016, Astronomische Nachrichten, 337, 105 

\bibitem[Sazonov et al.(2012)]{Sazonov12} Sazonov, S., Willner, S.~P., Goulding, A.~D., et al.\ 2012, \apj, 757, 181 

\bibitem[Schoenmakers et al.(2000)]{Schoenmakers00} Schoenmakers, A.~P., de Bruyn, A.~G., R{\"o}ttgering, H.~J.~A., et al.\ 2000, \mnras, 315, 371

\bibitem[Siemiginowska et al.(2008)]{Siemiginowska08} Siemiginowska, A., LaMassa, S., Aldcroft, T.~L., Bechtold, J., \& Elvis, M.\ 2008, \apj, 684, 811 

\bibitem[Siemiginowska et al.(2009)]{Siemiginowska09} Siemiginowska, A.\ 2009, AStronomische Nachrichten, 330, 264

\bibitem[Siemiginowska et al.(2016)]{Siemiginowska16} Siemiginowska, A., Sobolewska, M., Migliori, G., et al.\ 2016, \apj, 823, 57 

\bibitem[Singal et al.(2019)]{Singal19} Singal, J., Petrosian, V., Haider, J., et al.\ 2019, \apj, 877, 63

\bibitem[Sobolewska et al.(2019a)]{Sobolewska19a} Sobolewska, M., Siemiginowska, A., Guainazzi, M., et al.\ 2019a, \apj, 871, 71 

\bibitem[Sobolewska et al.(2019b)]{Sobolewska19b} Sobolewska, M., Siemiginowska, A., Guainazzi, M., et al.\ 2019b, \apj, 884, 166

\bibitem[Stanghellini et al.(1998)]{Stanghellini98} Stanghellini, C., O'Dea, C.~P., Dallacasa, D., et al.\ 1998, \aaps, 131, 303

\bibitem[Stawarz et al.(2008)]{Stawarz08} Stawarz, {\L}., Ostorero, L., Begelman, M.~C., et al.\ 2008, \apj, 680, 911 

\bibitem[Stern(2015)]{Stern15} Stern, D.\ 2015, \apj, 807, 129 

\bibitem[Stern et al.(2012)]{Stern12} Stern, D., Assef, R.~J., Benford, D.~J., et al.\ 2012, \apj, 753, 30 

\bibitem[Tadhunter(2016)]{Tadhunter16} Tadhunter, C.\ 2016, Astronomische Nachrichten, 337, 159 

\bibitem[Tadhunter et al.(2011)]{Tadhunter11} Tadhunter, C., Holt, J., Gonz{\'a}lez Delgado, R., et al.\ 2011, \mnras, 412, 960 

\bibitem[Taylor et al.(1998)]{Taylor98} Taylor, G.~B., Wrobel, J.~M., \& Vermeulen, R.~C.\ 1998, \apj, 498, 619

\bibitem[Tengstrand et al.(2009)]{Tengstrand09} Tengstrand, O., Guainazzi, M., Siemiginowska, A., et al.\ 2009, \aap, 501, 89 

\bibitem[Tingay et al.(1997)]{Tingay97} Tingay, S.~J., Jauncey, D.~L., Reynolds, J.~E., et al.\ 1997, \aj, 113, 2025

\bibitem[Tingay \& de Kool(2003)]{Tingay03} Tingay, S.~J., \& de Kool, M.\ 2003, \aj, 126, 723

\bibitem[Torresi et al.(2018)]{Torresi18} Torresi, E., Grandi, P., Capetti, A., Baldi, R.~D., \& Giovannini, G.\ 2018, \mnras, 476, 5535 

\bibitem[Ueda et al.(2005)]{Ueda05} Ueda, Y., Ishisaki, Y., Takahashi, T., Makishima, K., \& Ohashi, T.\ 2005, \apjs, 161, 185 

\bibitem[Van Velzen et al.(2015)]{VanVelzen15} van Velzen S., Falcke H., K{\"o}rding E. \ 2015, \mnras, 446, 2985

\bibitem[Vink et al.(2006)]{Vink06} Vink, J., Snellen, I., Mack, K.-H., \& Schilizzi, R.\ 2006, \mnras, 367, 928 

\bibitem[Wagner et al.(2016)]{Wagner16} Wagner, A.~Y., Bicknell, G.~V., Umemura, M., Sutherland, R.~S., \& Silk, J.\ 2016, Astronomische Nachrichten, 337, 167 

\bibitem[Watson et al.(2009)]{Watson09} Watson, M.~G., Schr{\"o}der, A.~C., Fyfe, D., et al.\ 2009, \aap, 493, 339 

\bibitem[Willett et al.(2010)]{Willett10} Willett, K.~W., Stocke, J.~T., Darling, J., \& Perlman, E.~S.\ 2010, \apj, 713, 1393 

\bibitem[W{\'o}jtowicz et al.(2019)]{Wojtowicz19} W{\'o}jtowicz, A., Stawarz, {\L}., Cheung, C.~C., et al.\ 2019, arXiv e-prints, arXiv:1911.01197

\bibitem[Wright et al.(2010)]{Wright10} Wright, E.~L., Eisenhardt, P.~R.~M., Mainzer, A.~K., et al.\ 2010, \aj, 140, 1868 


\end{thebibliography}
\end{document}